\documentclass{MLJCTemplate}
\usepackage[utf8]{inputenc}
\usepackage[english]{babel}
\usepackage{graphicx}
\graphicspath{{./images/}}
\usepackage[bookmarks=false]{hyperref}
\usepackage[backend=biber, sorting=none]{biblatex}
\usepackage{upquote}
\usepackage{csquotes}
\usepackage{amsmath}
\usepackage{amssymb}
\usepackage{mathtools}
\usepackage{comment}
\usepackage{bm}
\pdfoutput=1

\bibliography{biblio}
  {\begin{list}{}%
          {\setlength{\leftmargin}{#1}}%
          \item[]%
  }
  {\end{list}}

\usepackage{calrsfs}
\DeclareMathAlphabet{\pazocal}{OMS}{zplm}{m}{n}

\title{Optimizing Urban Mobility Restrictions: a Multi-Agent System (MAS) for SARS-CoV-2}

\author{\textbf{Simone Azeglio*\dag  \hspace{0.1cm} and \hspace{0.1cm} Matteo Fordiani*}}
\affil{}

\author{*Master Student, Department of Physics, University of Turin, \dag Visiting Student Researcher, University of Ottawa}
\affil{E-mail: simone.azeglio@edu.unito.it}

\thisvolume{XX}
\thismonth{October}
\thisyear{20XX}
\thisdoi{XX}
\thisstartpage{01}
\thisendpage{03}

\begin{document}

\maketitle 
\thispagestyle{thefirstpage}

\begin{center}
ABSTRACT 
\end{center} 

\vspace{-2mm}
\textbf{
Infectious epidemics can be simulated by employing dynamical processes as interactions on network structures. Here, we introduce techniques from the Multi-Agent System (MAS) domain in order to account for individual level characterization of societal dynamics for the SARS-CoV-2 pandemic. We hypothesize that a MAS model which considers rich spatial demographics, hourly mobility data and daily contagion information from the metropolitan area of Toronto can explain significant emerging behavior. To investigate this hypothesis we designed, with our modeling framework of choice, GAMA \cite{taillandier2019building}, an accurate environment which can be tuned to reproduce mobility and healthcare data, in our case coming from TomTom's API and Toronto's Open Data. We observed that some interesting contagion phenomena are directly influenced by mobility restrictions and curfew policies. We conclude that while our model is able to reproduce non-trivial emerging properties, large-scale simulation are needed to further investigate the role of different parameters. Finally, providing such an end-to-end model can be critical for policy-makers to compare their outcomes with past strategies in order to devise better plans for future measures. 
}

\section{\textbf{INTRODUCTION}}
In recent times, data availability  have created an enormous amount of new opportunities for investigators to improve on the traditional reporting of several diseases from international to national or
regional scale by studying variations in disease
occurrences at a local scale \cite{walter2000disease}. Such analysis may
include relevant health risk factor data
such as exposures to local sources of environmental pollution and the distribution of
locally varying social, economical and behavioral
factors. Modeling this type of event, also present new challenges from a statistical viewpoint: 
as the spatial scale of the investigation
becomes narrowed to a particular small zone, the reduced size of the population at risk leads to small numbers of infections and unstable estimates \cite{olsen1996cluster}.
From a pure statistical mechanical point of view, indeed, because of the small
population, such studies are more error prone than studies conducted over larger areas. At a broader scale,
purely local variations in data are likely
to largely cancel out (mean-field), whereas at the small-areas
scale, these variations could lead to serious
biases if not detected and corrected. 

People tend to cluster in
space in systematic ways that may be highly
predictive of disease risk: de facto creating communities. A typical example of segregation would be the one of people
of high socioeconomic status who tend to live near
others with high incomes and in areas with
better housing and schooling than those in
lower-income areas. Individuals with higher
incomes tend to have more favorable risk factor profiles (e.g., they are more likely to be
nonsmokers, take more leisure-time exercise) and as a consequence, have better health \cite{smith1996socioeconomic} \cite{smith1996socioeconomicII}.

We also note that an in-depth and individual based approach might be well suited to investigate how individuals interact with their environment and how
these interactions ultimately affect health. This could
address, for example, why people with higher
incomes take more leisure-time exercise. Is it
because they have a local environment more
enticing, have the financial resources to engage
in specific activities, have jobs that afford them
more leisure time, or pursue more leisure-time
activities for other reasons? Such questions
may have an important spatial component.
However, we see these as second-order issues
beyond the goal of this article.

Along with spatial epidemiological concerns, we also investigate the role of agents in reproducing human behaviours, by introducing \textit{Geographic Information Systems} (\textbf{GIS}) data as a basis in order to have a realistic environment. \textit{Agent-based modeling} (\textbf{ABM}), is a bottom-up approach which has the advanced capability of tracking the movement of a disease and the contacts between each individual in a social group located in a specific geographic area \cite{bagni2002comparison}, \cite{patlolla2004agent}. The potential that ABM possesses to model epidemic spread, has been used in epidemiology to study and track the movement of infected individuals and their contacts in a social system \cite{gordon2003simple}, \cite{dunham2005agent}.

Agent-based models allow interaction among individuals and are capable to overcome the limitations of different approaches such as classical epidemic models, permitting the study of specific spatial aspects in the spread of epidemics and addressing the stochastic nature of the epidemic process. Consisting of a population of individual actors or "agents", an environment, and a set of rules \cite{epstein1996growing}, actions in ABM take place through the agents, which are simple, self-contained programs that collect information from their surroundings and use it to determine how to act \cite{gilbert2005simulation}. Modeling an epidemic using an agent-based approach highlights the progression of a disease through each individual (thus populations become highly heterogeneous by health status during simulations), and tracks the contacts of each individual with others in the relevant social networks and geographical areas (e.g., co-workers, schoolmates). All the rules for movement at the individual level (\textit{e.g.} to and from a workplace or school) and for contacts and transmissions between people are explicit \cite{epstein2002toward}.

ABMs and their ability to produce emergent macro-effects from micro-rules have served as a cornerstone for the development of different methodological frameworks in epidemiology \cite{patlolla2004agent}. Epidemiological applications using ABM approach are mostly designed to allow epidemiological researchers to do a preliminary \textit{what if} analysis with the purpose of assessing a system's behaviour under diverse conditions and evaluating which alternative control policies to adopt in order to fight epidemics such as smallpox \cite{chen2004aligning} \cite{eidelson2004vir} \cite{carley2006biowar}. Although these models effectively track the progression of the disease through each individual, and track the contacts of each individual with others in the relevant system (social or natural), they need to add physical infrastructures such as road networks, and real geographic environments to account for more complex interactions among susceptible and infected individuals. 

The spread of human epidemics strongly relies on the structure of the underlying social network, and it has become clear that differently structured networks lead to different types of epidemiology \cite{keeling2011effects} \cite{chowell2003scaling}. By modeling the correlations between individuals, it is possible to understand the role of spatial heterogeneity in spreading dynamics. In our case, the use of real landscape structures and the integration of both geospatial data and mobility data, allowed us to represent the continuous environment where the discrete individuals interact. 

We now briefly consider the analytic
framework for carrying out spatial analyses and
the types of studies that we decided to undertake.
We then focus on a number of challenges that
face the practitioner of spatial epidemiology and multi agent systems, including issues of data availability and quality, exposure assessment, exposure mapping, and study design. 

\vspace{1mm}
\section{\textbf{DATA \& PREPROCESSING}}
Over the last few years we have been entering a new era of information technologies, and despite the evident potential of open data, the incessantly growing amounts of information being gathered by governments and industries is rarely accessible for scientists and researchers. 

There is a substantial vicious circle in the development of tools based on the vast ocean of data that we are facing nowadays: on one side there are lots of unstructured data that could be employed in scientific models for public service; on the other side, most of the times, researchers do not have at their disposal a sufficient amount of data to pinpoint their models.   

There is a certain amount of questions that need to be answered in order to circumvent such an issue and circumscribe its pros and cons: what kind of social and economical transformations has open data brought about, so far, and what transformations might it trigger in the future? How — and under which circumstances — has it been most effective? How have open-data practitioners mitigated risks (e.g. to privacy) while maximizing social good? \cite{verhulst2016open}

Open data are an open problem in science: especially  
when it comes to quantitative epidemic simulations, if your model aims at being effective for policy-makers at the spatial scale of a city, you would necessarily need data at a finer spatial scale - i.e. neighborhood. An analogous argument can be made for the temporal scale: if one's decide to include, let us say, traffic mobility dynamics, daily data are not enough to capture humans behaviour. 

Luckily, for some cities - the list is surprisingly very short - it is possible to gather open data at a fine spatial scale, both for zoning-by-laws and, in our case, for Sars-Cov-2 spread. Given the accessibility and huge amount of open data, we decided to build our model on the metropolitan area of Toronto - Canada. 

Here follows a brief overview of the different services (APIs) and datasets we employed to devise our model. 

\subsection{\textbf{Covid-19 Data - Toronto by Neighborhood}}

This dataset\footnote{Available at \href{https://open.toronto.ca/dataset/covid-19-cases-in-toronto}{https://open.toronto.ca/dataset/covid-19-cases-in-toronto/}/} contains demographic, geographic, and severity information  - at the spatial scale of each neighborhood - for all confirmed and probable cases reported to and managed by Toronto Public Health since the first case was reported in January 2020. This includes cases that are sporadic (occurring in the community) and outbreak-associated. The data are extracted from the provincial communicable disease reporting system (iPHIS) and Toronto's custom COVID-19 case management system (CORES) and combined for reporting. 

\graphicspath{{img/}}
\begin{figure}[htp]

\centering
\includegraphics[width=.5\textwidth]{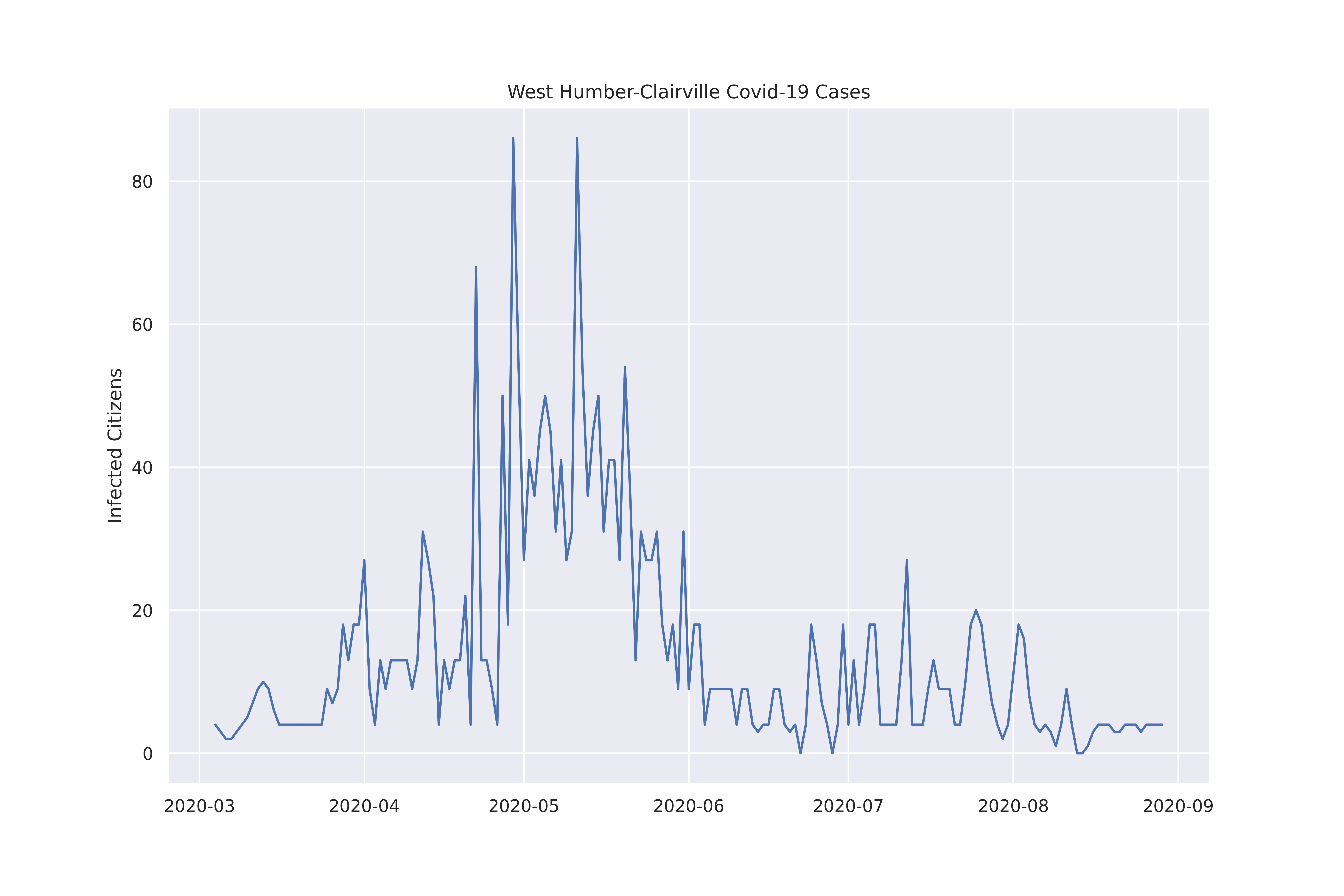}\hfill
\includegraphics[width=.5\textwidth]{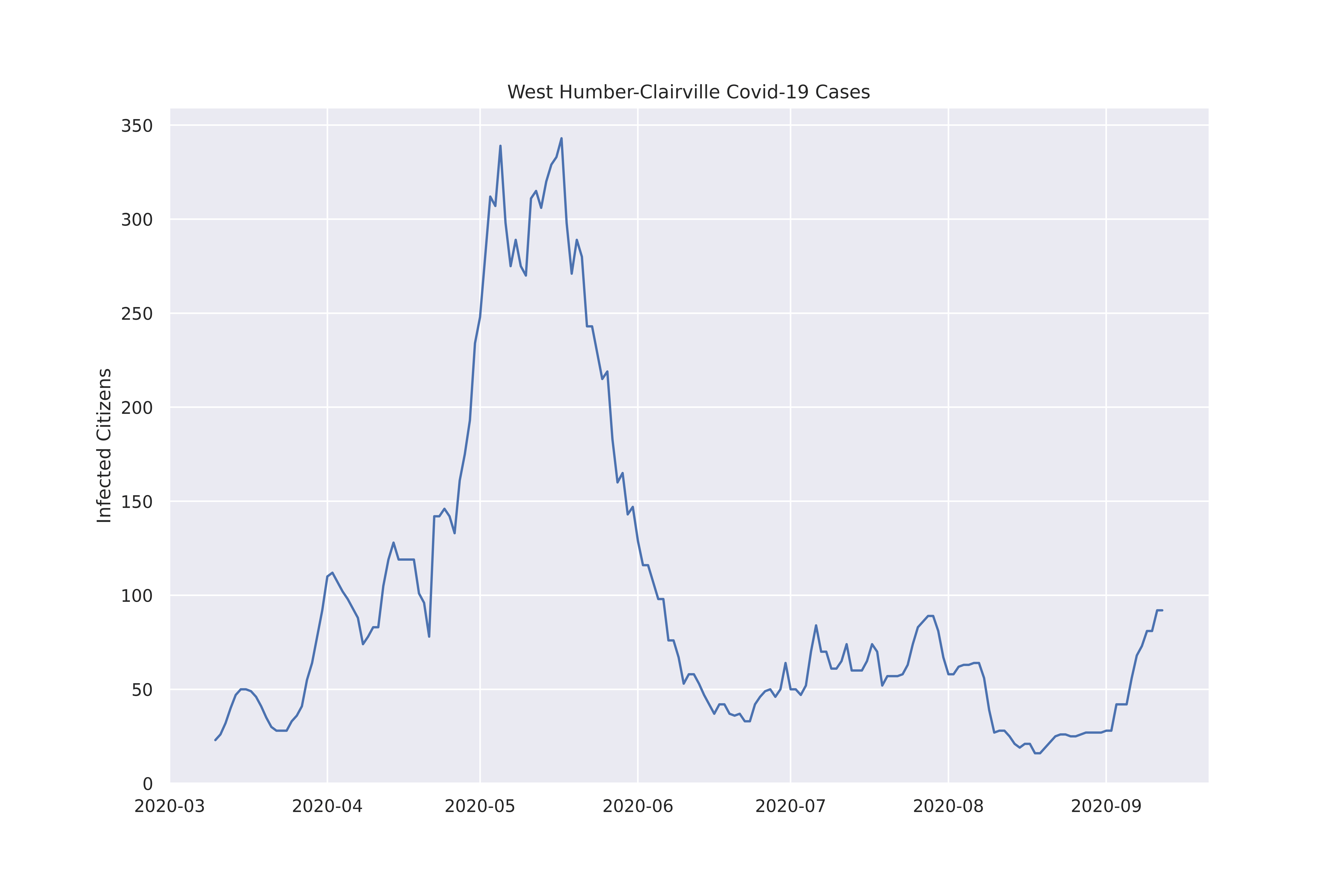}

\caption{\textit{left}: Covid Infection data for West-Humber-Clairville neighbourhood; \textit{right}: Rolling average (7 days) data for West-Humber-Clairville neighbourhood}
\label{fig:WHC_infected}

\end{figure}

\subsection{\textbf{TomTom Move API}}
TomTom, one of the leaders in location technologies, recently decided to develop an API service\footnote{Available at \href{https://developer.tomtom.com/traffic-api}{https://developer.tomtom.com/traffic-api}} in order to let their data be ready to use for different purposes. It is possible to select and download traffic mobility data for entire cities with a daily time resolution. With respect to the spatial scale, TomTom's API cover every existing street and road. Given this granularity, we decided to sample data from a pre-Covid time range, i.e. April 01-28 ,2019. 

In order to simulate the daily evolution of Sars-Cov-2 we needed some other data for extrapolating the hourly evolution of traffic mobility. For this purpose, we employed the historical TomTom \textit{Traffic Index}\footnote{Available at \href{https://www.tomtom.com/en\_gb/traffic-index/toronto-traffic/}{https://www.tomtom.com/en\_gb/traffic-index/toronto-traffic/}}- on the time range that we have previously selected - , which basically reports the hourly traffic congestion averaged over the whole area of interest.

\graphicspath{{img/}}
\begin{figure}[htp]
    \centering
    \includegraphics[width=0.5\textwidth]{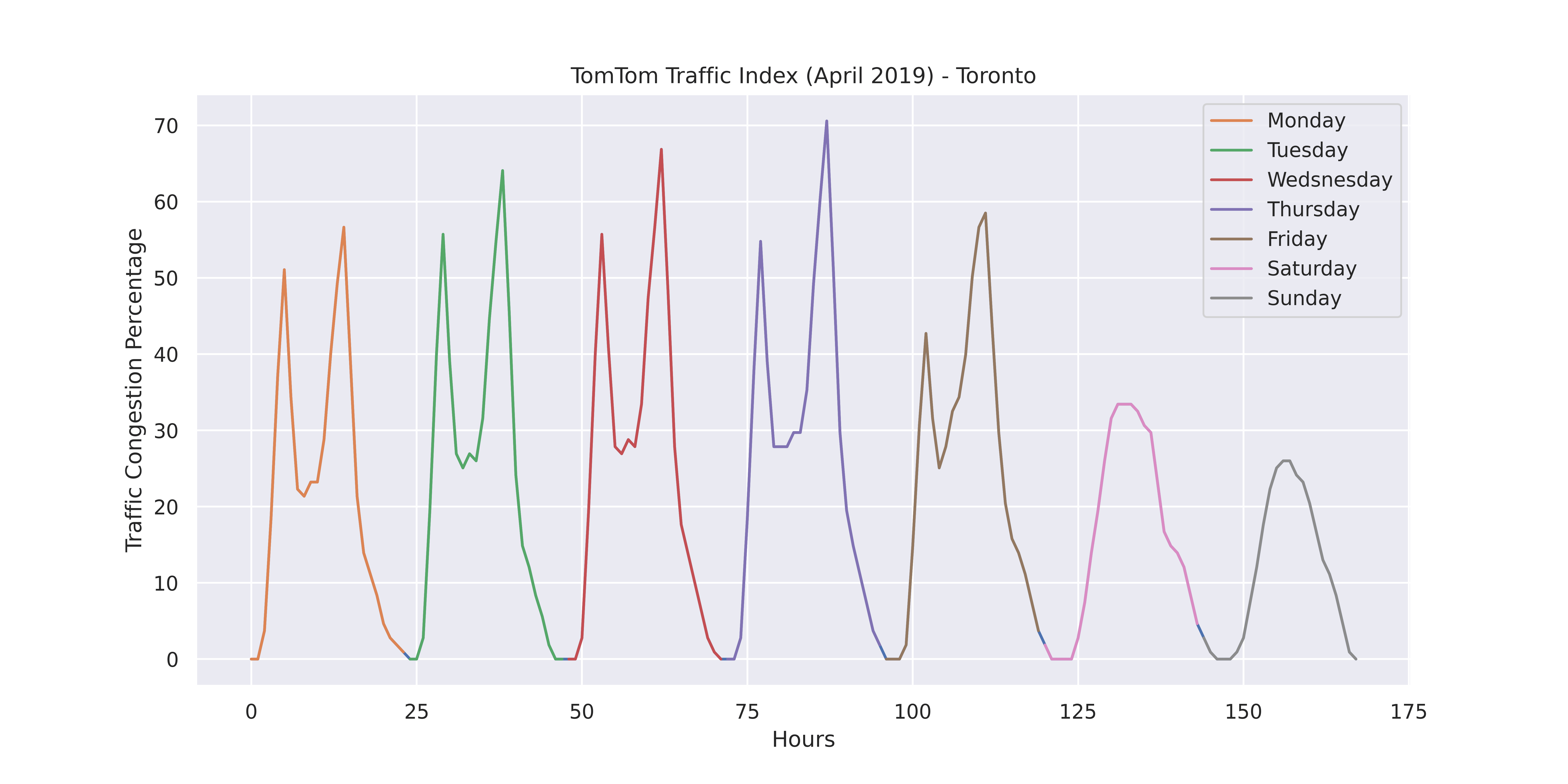}
    \caption{Hourly TomTom's Traffic Index - ToDo: fare meglio il grafico su notebook}
    \label{fig:traffic_congestion}
\end{figure}

\subsection{\textbf{OpenStreetMap (OSM) Layers}}

OpenStreetMap\footnote{Available at \href{https://www.openstreetmap.org/\#map=6/42.088/12.564}{https://www.openstreetmap.org/\#map=6/42.088/12.564}} is an astounding example of how open data - GIS in this case -  are crucial. OSM basically lets you export worldwide geodata for every map. In our case we employed  \textit{GeoFabrik}\footnote{Available at \href{https://www.geofabrik.de/data/}{https://www.geofabrik.de/data/}} and \textit{OSM Polygons}\footnote{Available at \href{http://polygons.openstreetmap.fr/}{http://polygons.openstreetmap.fr/}} to, respectively, download and extract the administrative area of Toronto and all its layers, such as Railways, Transport Lines (Subways, Bus routes), Airports; which turned out to be useful in constructing our MAS model.

\subsection{\textbf{Toronto's Neighborhood Profiles}}

This dataset contains Neighbourhood Profiles which are based on data collected by Statistics Canada in its 2016 Census of Population. Data is gathered from the Census Profile as well as a number of other tables, including Core Housing Need, living arrangements and income sources. 

\subsection{\textbf{3D Massing Data}}

This dataset\footnote{Available at \href{https://ckan0.cf.opendata.inter.prod-toronto.ca/tl/dataset/3d-massing}{https://ckan0.cf.opendata.inter.prod-toronto.ca/tl/dataset/3d-massing}} is part of the impactful project of the City of Toronto in creating city-wide open source data in accessible formats. By exploring such a way of sharing information, a direct consequence is an improvement in the planning process, from different perspectives. In our case we are interested in this dataset since it provides a variety of 3D digital models, readily available to the public. 

Essentially, it provides us with 3D models of buildings, which might be useful in re-creating the population distribution in a realistic manner, by giving us the chance to roughly reproduce apartments and families dispositions.

\subsection{\textbf{Zoning by Laws Data}}

This dataset\footnote{Available at \href{https://ckan0.cf.opendata.inter.prod-toronto.ca/dataset/zoning-by-law}{https://ckan0.cf.opendata.inter.prod-toronto.ca/dataset/zoning-by-law}} contains \textit{ESRI shapefiles} that are part of the \textit{Zoning By-law 569-2013}. Basically data are divided in zones which are assigned to a specific category, e.g. Residential, Open Space, Commercial, Institutional and many more. 

\subsection{\textbf{Apple Mobility Trends Reports}}
Since TomTom's mobility data are extracted from a pre-Covid-19 period, it might be essential to find a way to constrain eventual lockdown or partial-lockdown strategies by affecting mobility. Apple publishes , with the highest level of privacy, daily reports which reflect requests for directions on Apple Maps. 
In particular, since some days were missing, we opted for a \textit{geometric progression} imputing, in order to preserve the general trend of the time series. 

\begin{figure}[htp]
    \centering
    \includegraphics[width=0.5\textwidth]{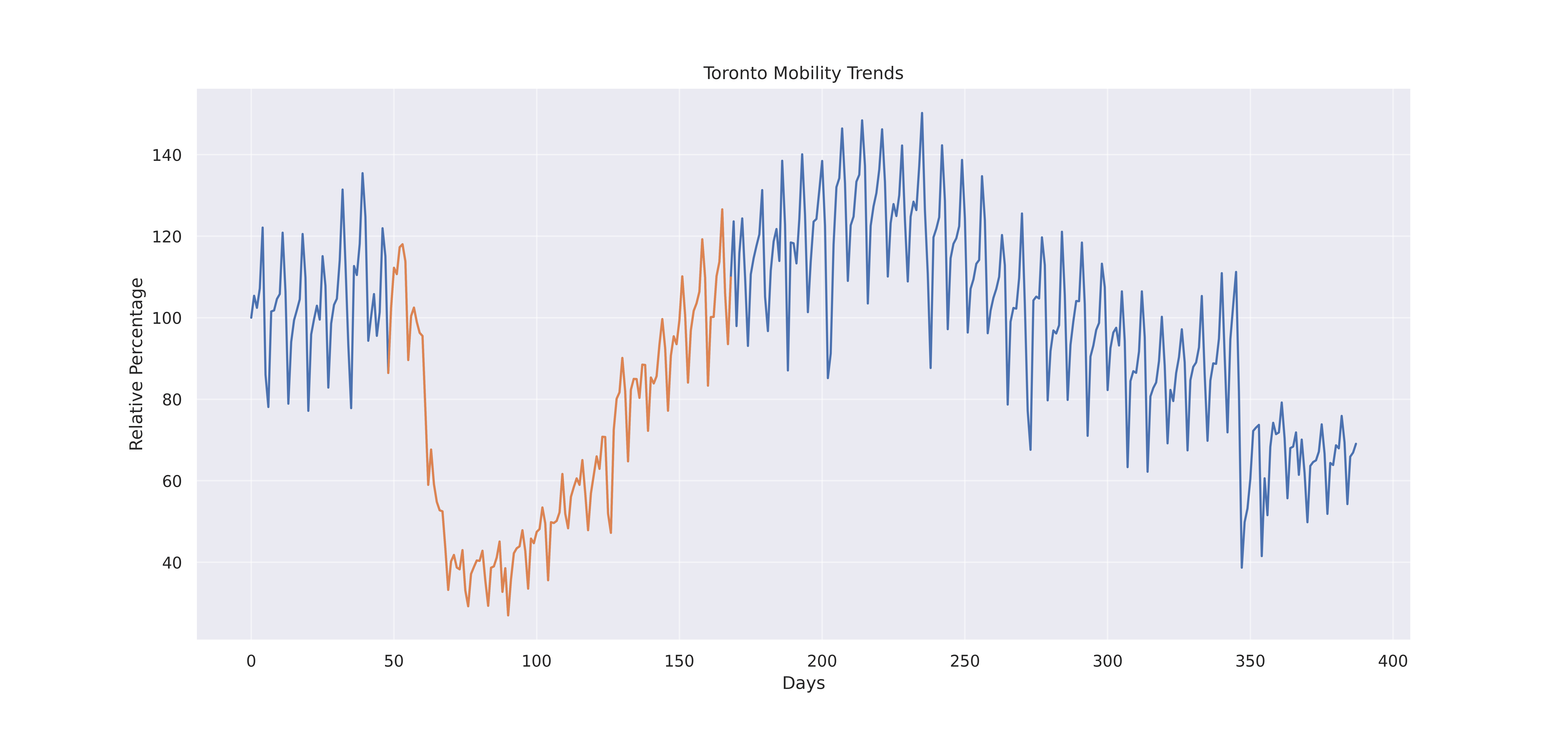}
    \caption{Apple Mobility Trends, in \textit{blue} the whole time series, day 0 corresponds to 01-13-2020, while in \textit{orange}, the simulated time range (corresponding to the first wave), 120 days starting from 03-01-2020}
    \label{fig:mobility}
\end{figure}

\section{\textbf{METHODS}}

\subsection{\textbf{GIS Data Preprocessing}}

In order to build a realistic environment for our agents, we needed to merge two GIS datasets, namely: \textit{3D Massing} and \textit{Zoning by Laws}. By using \textit{geopandas}\cite{kelsey_jordahl_2020_3946761} Python library, we have been able to, firstly match coordinates projection (epsg : 4326) and secondly to preprocess both datasets. The last step in this process has been to intersect zoning polygons with buildings, in order to extract each building category.

In this way we obtained an environment where each building is assigned to some specific category, de facto making an agent's life easier in choosing where to head to. 

\graphicspath{{img/}}
\begin{figure}[htp]
    \centering
    \includegraphics[width = 0.5\textwidth]{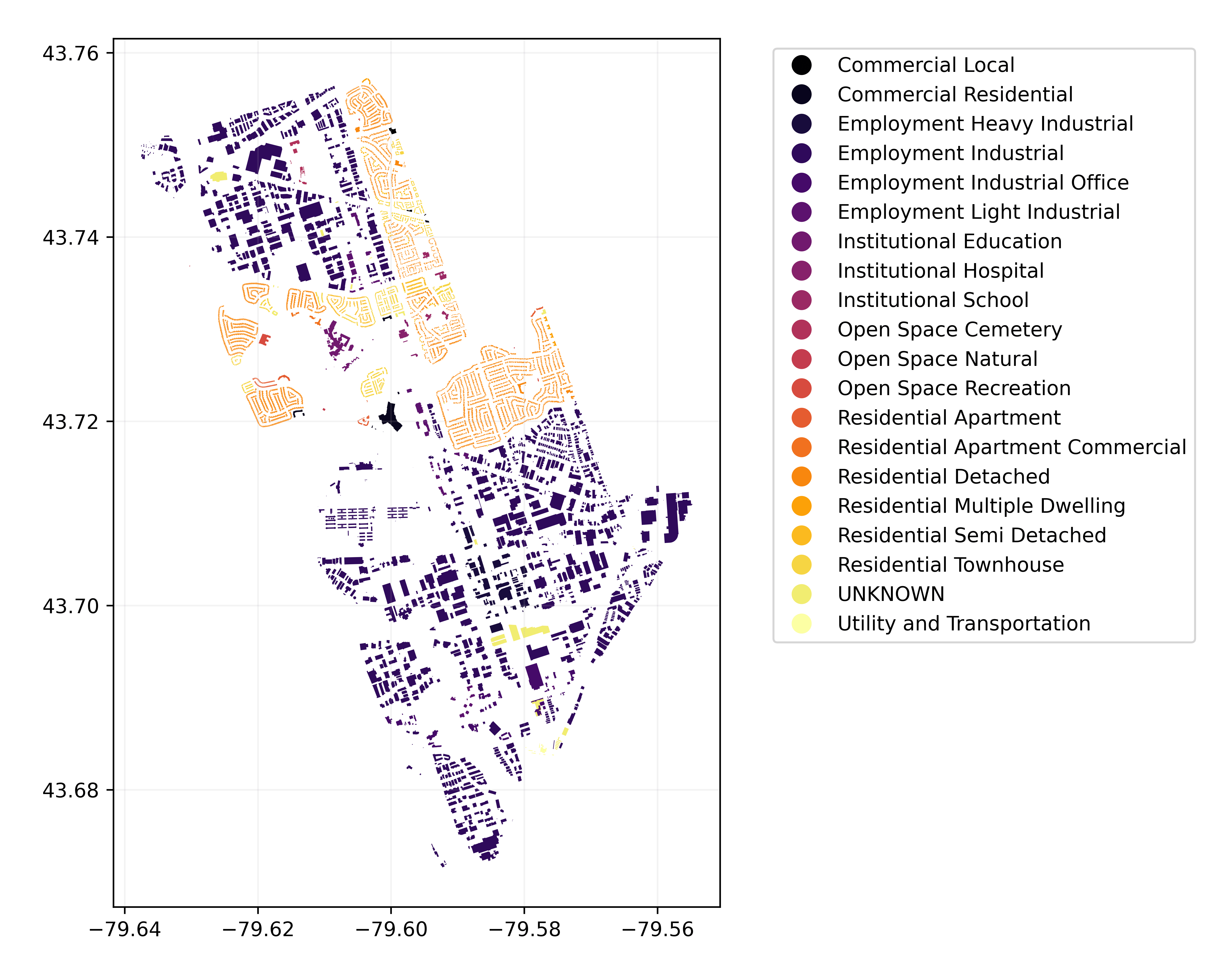}
    \caption{Preprocessed GIS Data, Zoning by law by building for West-Humber-Clairville neighborhood}
    \label{fig:Toronto_Zones}
\end{figure}

As a proof of concept, we decided to investigate the application of our model on a restricted area environment. Instead of employing the whole city of Toronto, which is definitely a medium-large sized city, we decided to focus our attention on a single neighborhood: \textit{West Humber-Clairville}. Again, by employing \textit{geopandas} it is not difficult to retrieve a specific neighborhood and extract its perimeter for further processing of demographic Data.

\subsection{\textbf{Demographic Data Preprocessing}}

\textit{Neighborhood Profiles} have been useful to determine the population for each specific neighborhood, as well as the age distribution. Luckily there were information regarding the distribution of households, but up to 4 members. To build a more realistic environment, we decided to empirically estimate households with 5 and more members: by fitting an exponential function on the number of households from 1 to 5, we have been able to impute the missing data.

\begin{figure}[htp]

\centering
\includegraphics[width=0.5\textwidth]{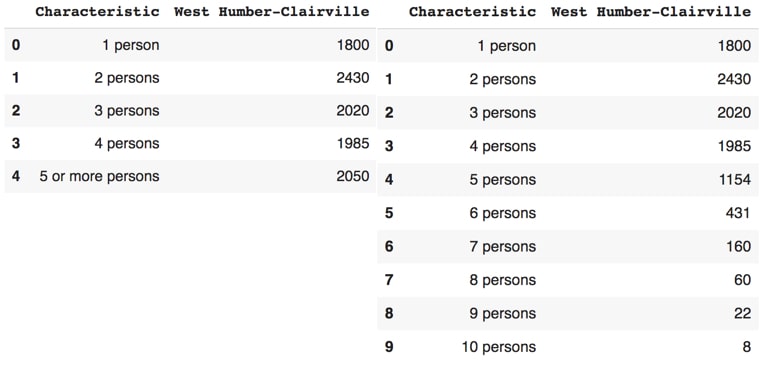}

\caption{\textit{Left}: Original households statistics for West-Humber-Clairville neighborhood; \textit{right}: Households statistics up to 10 persons, after exponential fit}
\label{fig:WHC_Houselholds}

\end{figure}


\subsection{\textbf{Traffic Data Preprocessing}}

Roads exhibit a network structure, and traffic can be envisioned as a time evolving process on a network. 
The starting point has been to employ OSM Layers and specifically the road layer as the skeleton or static network of our model.

In order to replicate realistic mobility patterns, we decided to employ TomTom API's data as a building block for outlining a weighted network approach. One fundamental feature, named \textit{sampleSize}, has been employed as a criterion for extracting reliable weights. The \textit{sampleSize} feature is basically the daily number of vehicles going down a specific street. We needed to integrate the \textit{Traffic Index} so that to get an hourly weighted network. By doing that, the result wasn't significantly different: daily weights and hourly weights were pretty much the same (up to the 5th decimal), that's why we decided to stick to the first one and we assumed that weights are static during the day. 

Having such a weighted network allows us to extend our current model to include public transports with realistic travel time and thus realistic infection conditions. 

\begin{figure}[htp]
    \centering
    \includegraphics[width = 0.5\textwidth]{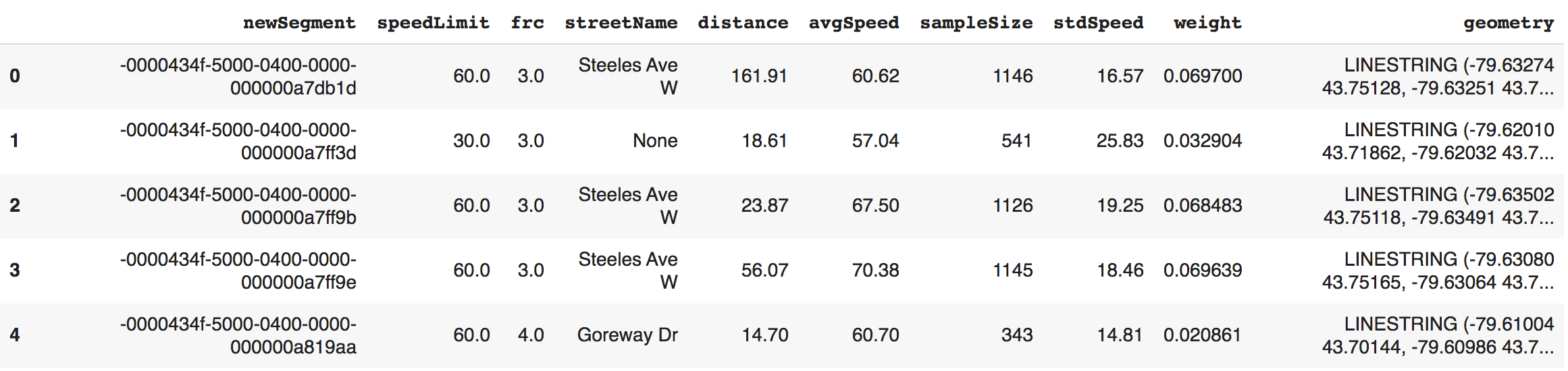}
    \caption{Traffic Data from TomTom Traffic API, preprocessed with \textit{geopandas}}
    \label{fig:Traffic_Preprocessing}
\end{figure}


\begin{figure}[htp]

\centering
\includegraphics[width=.5\textwidth]{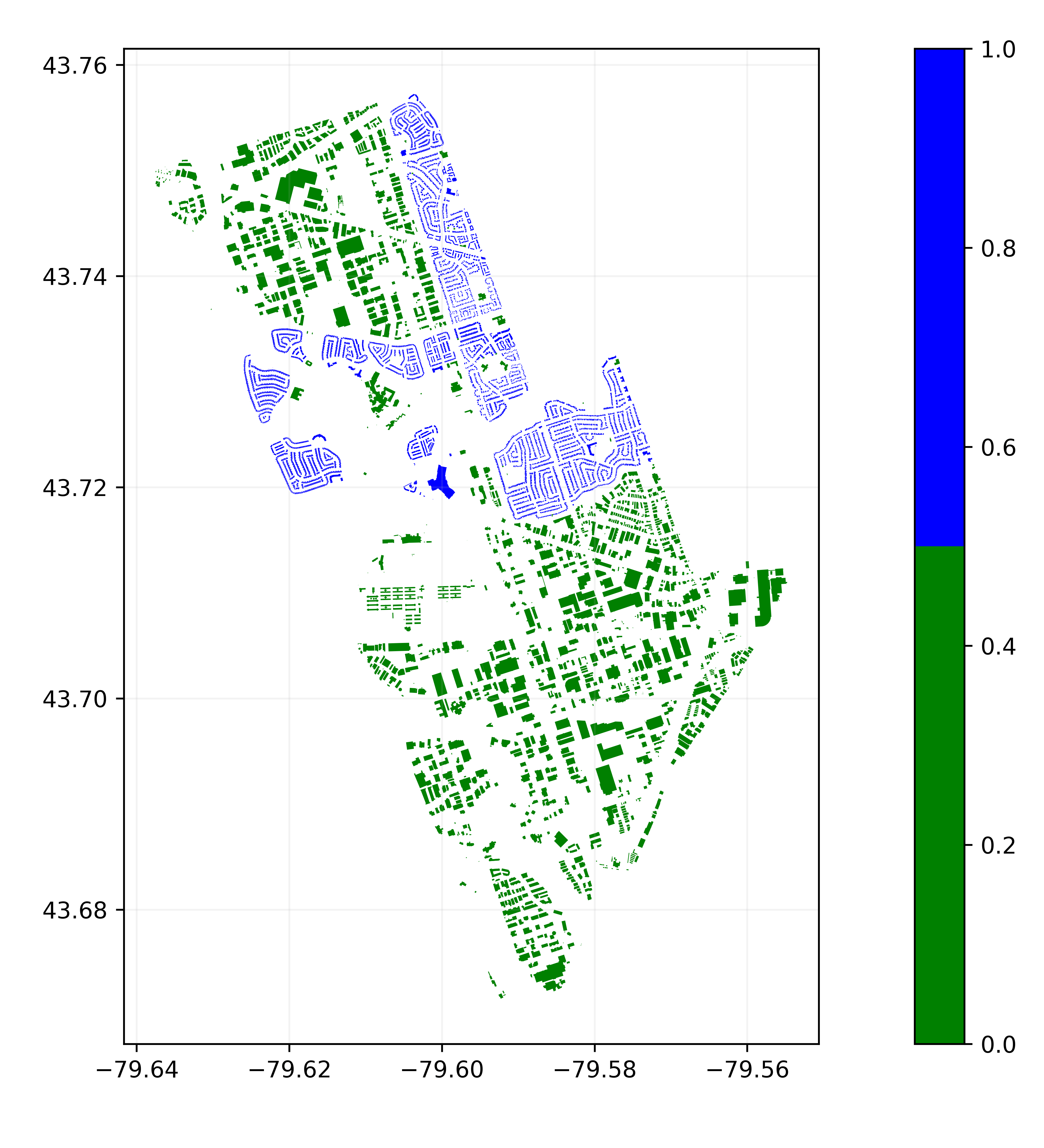}\hfill
\includegraphics[width=.5\textwidth]{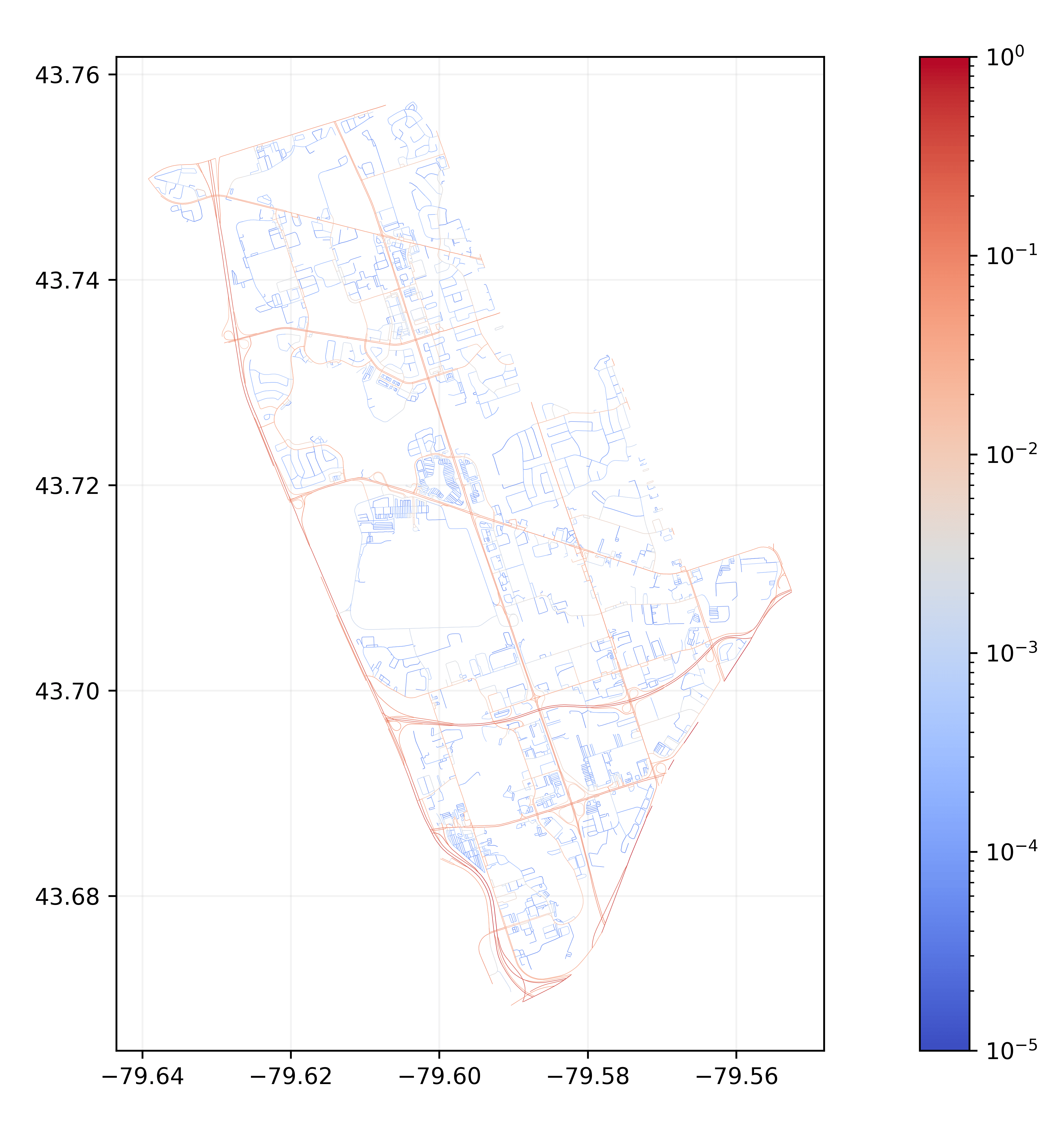}

\caption{\textit{left}: Demographic Data for West-Humber-Clairville neighborhood, \textit{colors}: blue = living place; green = non-living place ; \textit{right}: Weighted Road Network for West-Humber-Clairville neighborhood, \textit{colors} = $log(weight)$}
\label{fig:double_toronto}

\end{figure}

\subsection{\textbf{Multiplex Networks Approach}}

The term \textit{multiplexity}, whether it denotes "the co-existence of different normative elements in a social relationship" - as stated by Max Gluckman in \cite{gluckman1962essays} -  or the co-existence and overlap of different activities in relationships, where, in this specific case, social relationships are intended as media for the transmission of of different types of information, is definitely a framework for measuring various bases of interaction in a networked system.  

We are interested in separating locus-specific infection dynamics, with the intention to further extend this model, in the near future, to support contact-tracing for a more efficient mitigation plan. 

For the sake of clarity we give a more formal introduction to Multiplex Networks by exploiting Graph Theory. 
A networked system \textbf{N} can be represented by a graph. A graph is a tuple $G(V,E)$, where $V$ is a set of nodes and $E \subseteq V \times V$ is a set of edges, where each edge connects a pair of nodes. In this instance, nodes represents buildings where a person goes to, while edges are essentially social interactions. If node $u$ and node $v$ are connected by an edge in $G$, i.e. $(u,v) \in E$, they are said to be adjacent. It is possible to define a graph by employing the concept of adjacency, by means of an adjacency matrix $A_{adj}$. 

In order to represent a network in which different types of relations exist between the components — a multiplex network — we must introduce the notion of \textit{layer}. In our model layers correspond to buildings' (zoning by laws) classes - e.g. Residential buildings, Worship buildings. Let $L = \{1, . . . , m\}$ be an index set, which we call the \textit{layer set}, where $m$ is the number of different layers, i.e. the number of different kind of relations among nodes. 

Now, consider our set of nodes $V$  and let $G_P = (V , L, P )$ be a binary relation, where $P \subseteq V \times L$. The statement $(u, \alpha) \in P$, with $u \in V$, and $\alpha \in L$, is read node $u$ participates in layer $\alpha$.  In other words, we are attaching labels to nodes that specify in which type of relations (layers) the considered node takes part in.

Finally, consider the graph $G_C$ on $P$ in which there is an edge between two node-layer pairs $(u, \alpha)$ and $(v, \beta)$ if and only if $u = v$; that is, when the two edges in the graph $G_P$ are incident on the same node $u \in V$ , i.e. the two node-layer pairs represent the same node in different layers. We call $G_C(P , E_C)$ the coupling graph. It is easy to realize that the coupling graph is formed by $n =| P |$ disconnected components that are either complete graphs or isolated nodes. Each component is formed by all the representatives of a node in different layers, and we call the components of $G_C$ supra-nodes.
We are now in the position to say that a multiplex network is represented by the quadruple $M = (V,L,P,M)$:

\begin{itemize}
    \item the node set V;
    \item the layer set L ;
    \item the participation graph $G_P$; 
    \item the layer-graphs M 
\end{itemize}

Next, consider the union of all the layer-graphs, i.e., $G_l = \cup_\alpha G_\alpha$ We call such a graph the intra-layer graph. Note that, if each layer-graph is connected, this graph is formed by $m$ disconnected components, one for each layer-graph.
Finally,we can define the graph $G_M =G_l \cup G_C$,which we call the supra-graph. $G_M$ is a synthetic representation of a multiplex network.

It follows that, in this way, we can define an adjacency matrix for each layer, namely : \textit{layer-adjacency matrix}. On the other side, we can define another adjacency matrix for supra-graphs : the \textit{supra-adjacency matrix}.

The goal of this network approach, integrated in a multi-agent model, is to offer a static benchmark for observed contagion paths: recent results by L. Carpi \textit{et al.} \cite{carpi2019assessing} \cite{oliveira2020multiplex} on diversity in multiplex networks suggest that layers adding diverse connectivity to the multiplex (i.e. inter-layer connectivity) account for a multiplex network's efficiency. In this context "efficiency" roughly translates to fairly short paths between any couple of nodes, using nodes that are represented in two or more layers as gateways; in our model this behavior is represented by a contagion that explodes in a work environment, is passed onto family members and thus reaches their workplaces as well. 

If, by means of extensive computation and real-world studies, we can reasonably prove that the dynamical evolution of an epidemic is strongly tied to measures of diversity and efficiency in a multiplex network, effective restrictive measures could be implemented with surgical precision, based on network structures alone. This goes beyond the scope of this paper, nonetheless we have started feasibility studies on the matter and for visualization purposes we're including a snapshot of the multiplex network underlying our epidemic modeling: we decided to focus on a \textit{Egocentric Network}.  Egocentric networks are local networks with one central node, known as the \textit{Ego}. The network is based off the ego and all other nodes directly connected to the ego are called \textit{alters}.

An Ego is the focal point of the network during data collection and analysis and are ‘surrounded’ by alters. 
Here you can see the supra-adjacency matrix for our ego network of choice, on one side, while on the other the multiplex network. 

\begin{figure}[htp]

\centering
\includegraphics[width=.5\textwidth]{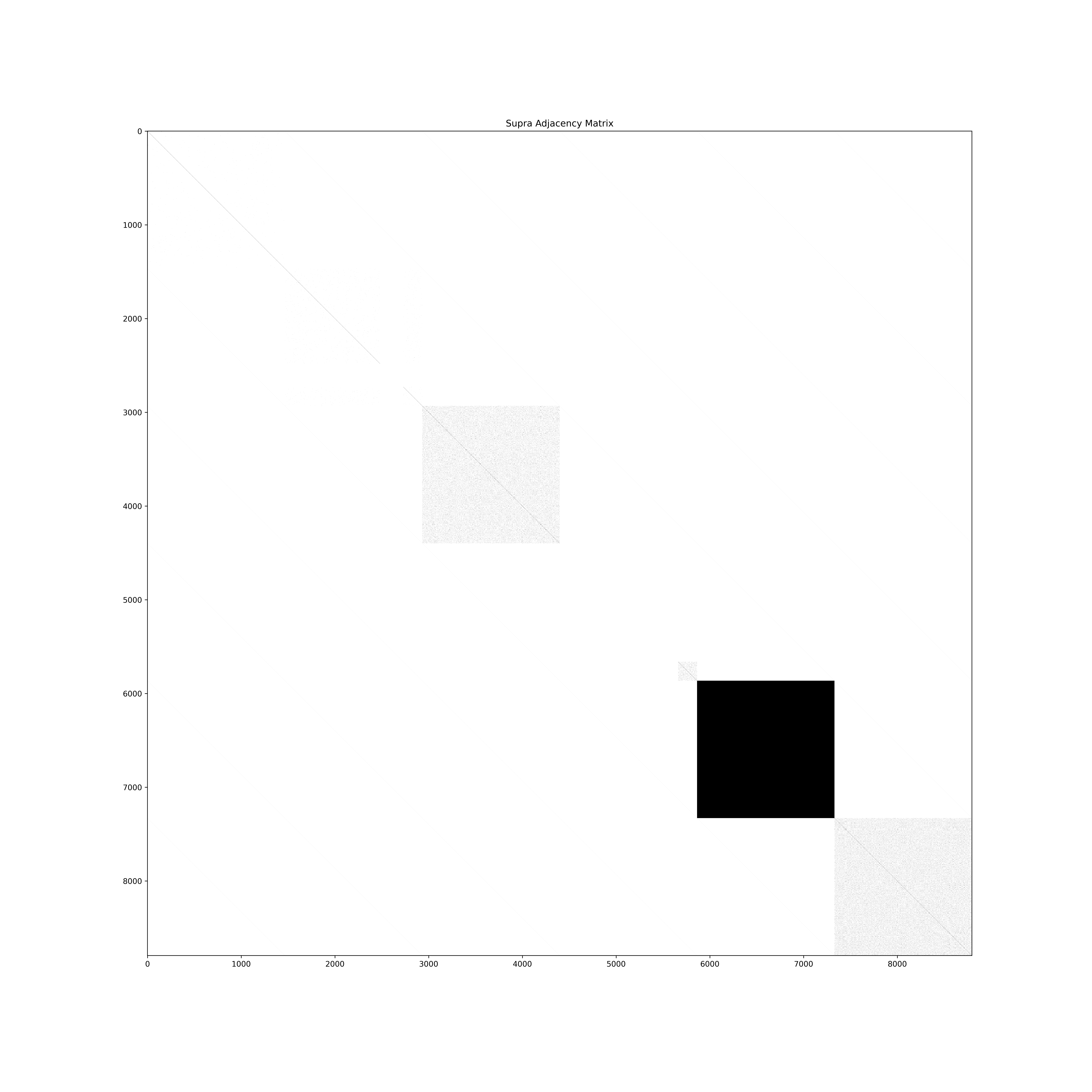}\hfill
\includegraphics[width=.5\textwidth]{img/Multiplex_Toronto_Ego.jpg}

\caption{\textit{top}: supra-adjacency matrix, each squared submatrix represents the adjacency matrix of a specific layer: \textit{living}, \textit{working}, \textit{worship}, \textit{school}, \textit{stuff1}, \textit{stuff2};
\textit{bottom}: Multiplex Ego Network,each layer from bottom to top:  \textit{living}, \textit{working}, \textit{worship}, \textit{school}, \textit{stuff1}, \textit{stuff2}.}

\label{fig:multiplex_toronto}

\end{figure}

\section{\textbf{MODEL OUTLINE}}
\label{headings}

\subsection{\textbf{Environment Modeling}}
As detailed in previous sections, the environment is made up of two species of agents: \textit{Roads} and \textit{Buildings}. \textit{People} move through \textit{Roads} to reach \textit{Buildings}.

\textit{Roads} are nothing more than weighted connectors between \textit{Buildings}: their epidemiological effect, so far, is reducing the overall time \textit{People} spent in \textit{Buildings}: we model real-world speed and, weighting the networks based on real traffic (\textit{People} prefer greater-weight paths for same endpoints), congestion effects ensue, leading to slow fluxes of \textit{People} from their \textit{Homes} to their \textit{Workplaces} and vice-versa.

\textit{Buildings} are at the core of our epidemic: we assume that the main factor determining social network structure and risk of infection is the kind of place we're looking at: \textit{Homes} will have more high-risk contacts than \textit{Workplaces}, but in \textit{Workplaces} teams are usually tightly connected through the chain of command, while in most \textit{Residential Apartments} of a large city (as Toronto is) families will hardly interact with each other. These are just assumptions, which need validation from sociological and demographic studies. 
One of the goals of this paper is to outline how, with cooperation between sociologists, epidemiologists, policy-makers and computational scientists, a highly representative model can be set-up and used to test our assumptions on epidemics spreading and to optimize restriction policies, preserving public health with the least damage to local economies.

\begin{figure}[htp]

\centering
\includegraphics[width = 0.5\textwidth]{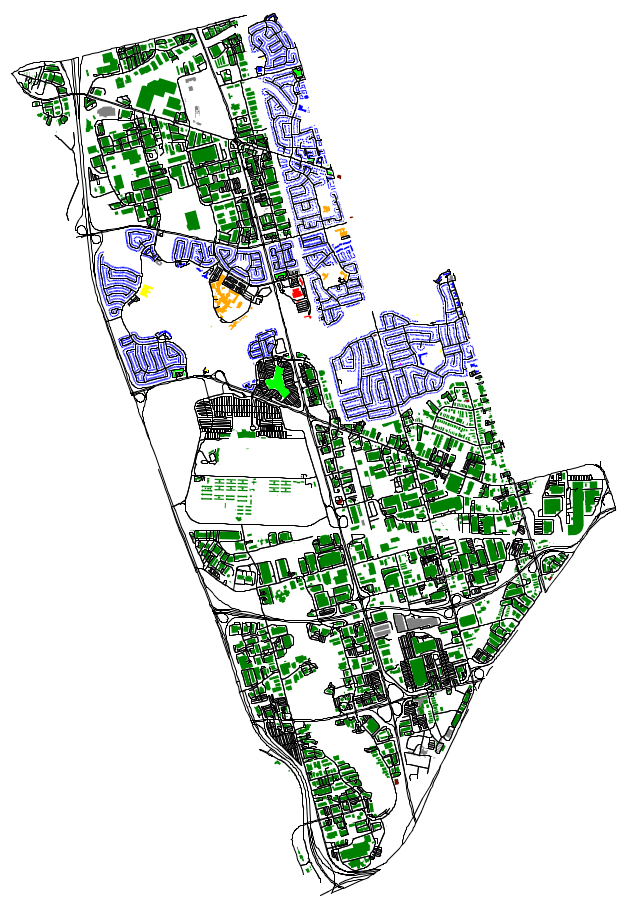}\hfill

\caption{Map of West-Humber-Clairville (Toronto neighbourhood) generated in GAMA, color-coded following zoning by laws.}
\label{fig:MAP_colored}

\end{figure}
\subsection{\textbf{Basic Skills: Movement}}

Only \textit{People} possess the ability to move. In GAMA movement is achieved by setting target \textit{Buildings}.
With respect to movement we identify a binary state-space: \textit{People} can be \textit{Resting} or \textit{Working}.
While an agent's target is wiped at destination, its \textit{Objective} is stored as a property, thus allowing us to model realistic schedules and differentiate by age class. 
We introduce stochastic visits to shops and open-spaces in the agent's free time. To avoid non-realistic movements a \textit{staying coefficient} multiplies the probability to take an unscheduled trip to the shop (or to an open-space), bringing it towards zero at lunch, dinner and night-time.

We model part-time and full-time workers by assigning randomly drawn start and end hours for work activities. We draw from a uniform distribution over the typical workday (start is in $[6AM, 9AM]$, end is in $[16PM,20PM]$).

\begin{figure}[htp]

\centering
\includegraphics[width = 0.5\textwidth]{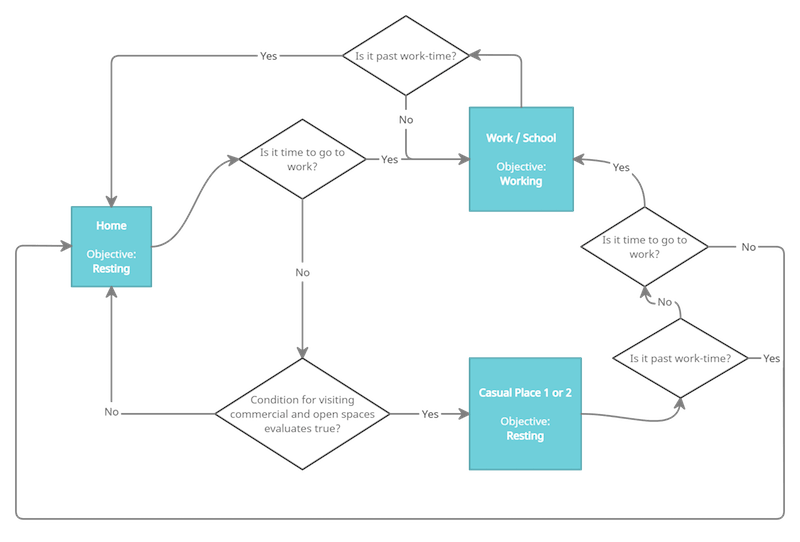}

\caption{Movement flow-chart for \textit{Juveniles} and \textit{Adults}}
\label{fig:MOVEMENT}

\end{figure}

In Figure \ref{fig:MOVEMENT} we portray the movement flow-chart for \textit{Adults} and \textit{Juveniles}. \textit{Seniors} are cut out of the work loop altogether, they are in a permanent \textit{Resting} state, so their chance to take unscheduled trips to stores and parks is greater.

\subsection{\textbf{Social Networks Modeling}}
To model social networks commonly found and studied in offices, schools and communities at large, we adopt a simple generative approach: 
\begin{itemize}
    \item Groups inside a building are considered cliques: they are complete graphs
    \item At initialization, every \textit{Person} is assigned a \textit{Residential Building}, its home. Depending on its age class, it is assigned a \textit{Workplace}, a \textit{School} and two \textit{Leisure Locations}.
    \item  On the first cycle every building iterates the \textit{People} vector to find its \textit{Inhabitants}. At this stage we're not discriminating between workers and visitors: \textit{People} will be able to infect or be infected only if if they're actually present in the \textit{Building}.
    \item The \textit{Building} splits its inhabitants into \textit{Groups}, which size is drawn from a uniform distribution between 2 and 7 (included). There's a 10\% chance a person in the \textit{Group} will be included in the next \textit{Group} as a link, thus only adjacent \textit{Groups} are linked: this allows a daisy-chain of infections in a single time-step. All but \textit{Residential Buildings} follow these rules: any kind of \textit{Residential Building} spits its inhabitants into isolated cliques which size is drawn from a uniform distribution between 1 and 8, representing families.
\end{itemize}

\begin{figure}[htp]

\centering
\includegraphics[width = 0.5\textwidth]{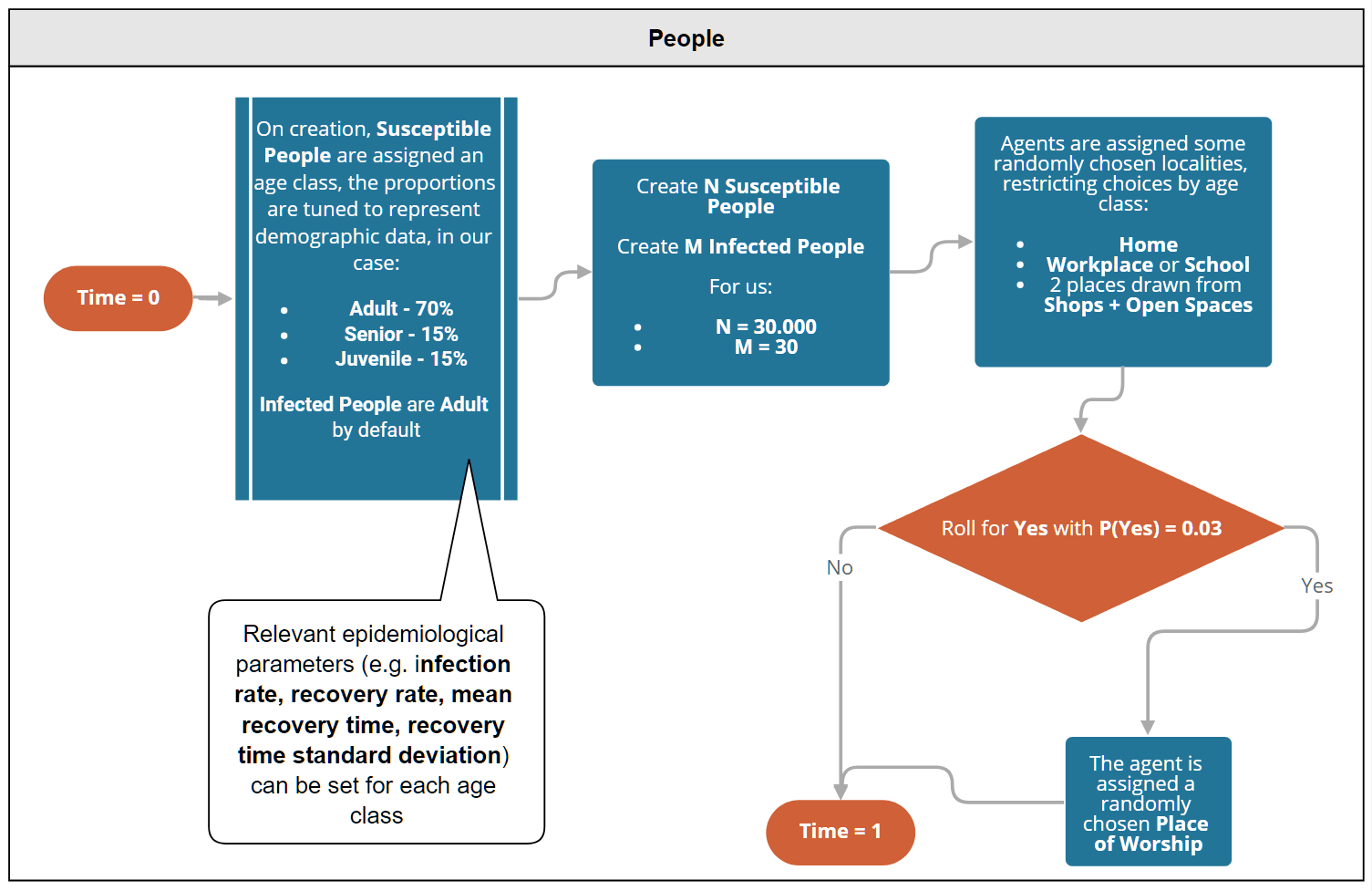}

\caption{Initialization flow-chart for \textit{People}}
\label{fig:PEOPLE_INIT}

\end{figure}

\begin{figure}[htp]

\centering
\includegraphics[width = 0.5\textwidth]{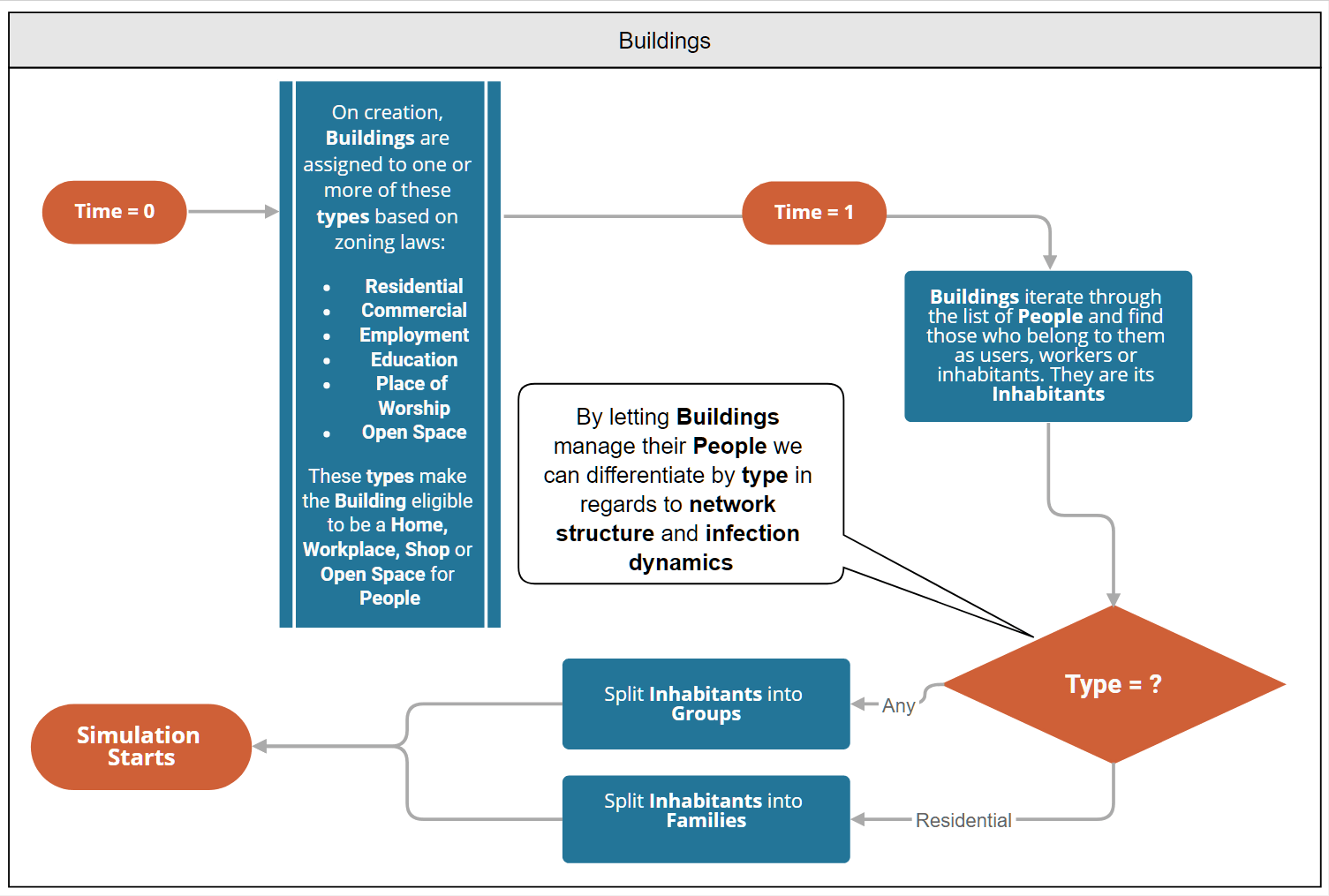}

\caption{Initialization flow-chart for \textit{Buildings}}
\label{fig:BUILDINGS_INIT}

\end{figure}

\begin{figure}[htp]

\centering
\includegraphics[width = 0.5\textwidth]{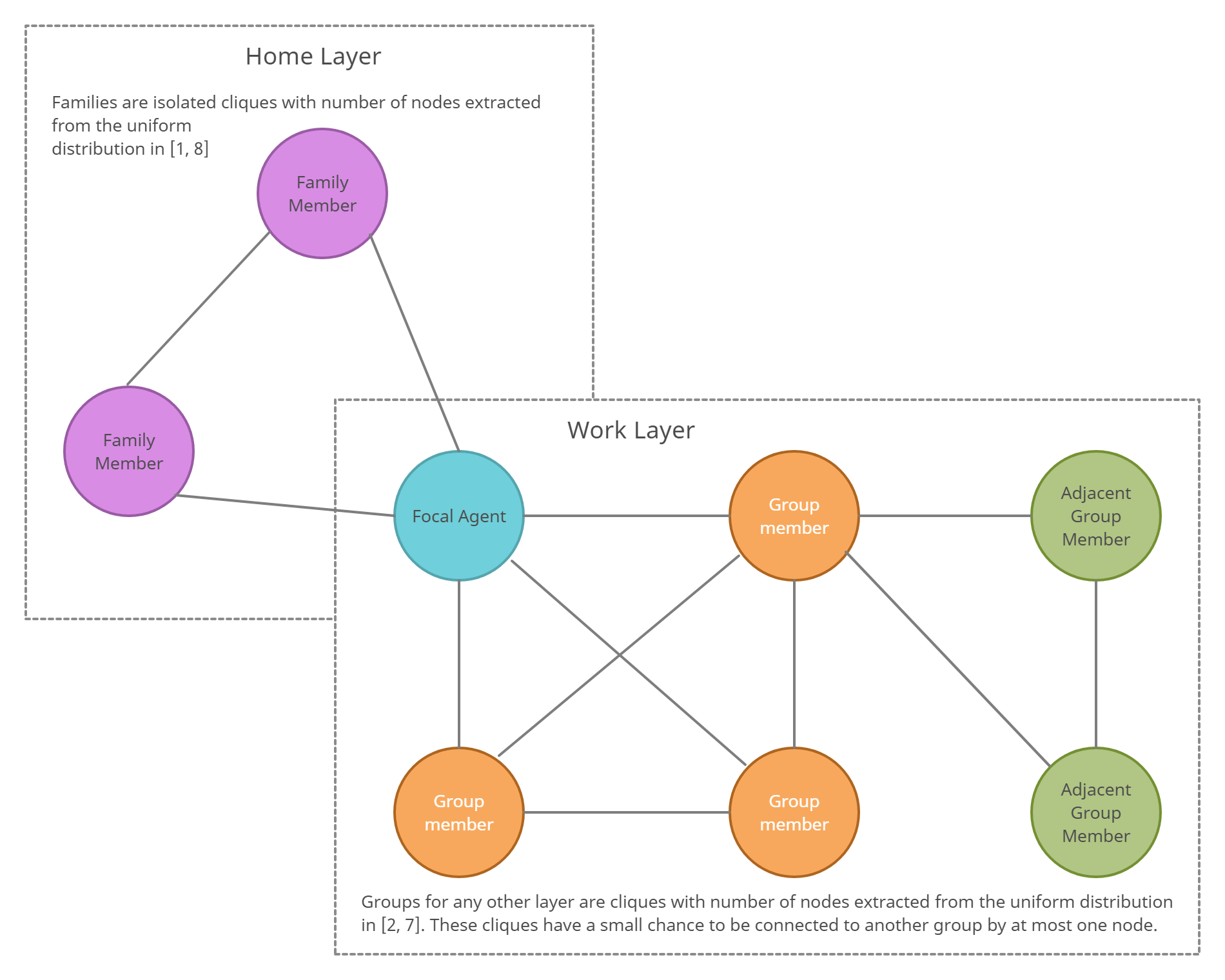}

\caption{Social network of a focal agent}
\label{fig:PEOPLE_NETWORK}

\end{figure}

Figures \ref{fig:PEOPLE_INIT} and \ref{fig:BUILDINGS_INIT} detail flow-charts for our initialization process, while Figure \ref{fig:PEOPLE_NETWORK} shows the graph structure of our social networks for a focal agent.

\subsection{\textbf{Epidemic modeling}}
Given the agent-based nature of our model, we decided to model micro-interactions using minimal assumptions on widely employed compartmental models:

\begin{itemize}
    \item Contagion paths are due to network structure alone.
    We neglect environmental contagion and assume a systematic error on our probability of contagion due to one high-risk contact. 
    \item The social function of a building uniquely determines the underlying social network connecting its users. Social groups are assumed to be cliques.
    \item The chance to be infected by family members is higher than the chance to be infected from colleagues or acquaintances.
    \item No contagion takes place on roads or on public transport as no persistent social network is there to be found.
    \item Intragroup contagion, at each step, follows a voter model, where the probability of being influenced equals the probability of contagion due to one high-risk contact. We assume at most two high-risk contacts can happen per day between any pair of individuals in the same group.
    \item We consider recovering and developing immunity as the same event. 
    \item The fraction of infected that eventually recovers is considered a characteristic of the system under study to be measured, thus we do not randomize it or try to explore its values. Instead, we draw the time-distribution of immunization events from a normal distribution (i.e. the minimum time that must pass between infection and a roll for immunization).
    \item Death happens without re-insertion.
\end{itemize}

We employ SIR compartments, where the internal dynamics of the \textit{I} compartment are broken down into \textit{Infected} and \textit{Hospitalized}. \textit{Hospitalized} people are counted in the \textit{I} compartment but are assumed to be isolated and cannot infect. After 2 weeks they can either die or develop immunity and get back to their business.

We want to emphasize that compartmental models can get as rich as one pleases, but our focus here is on method and simulation architecture. Crisis-prevention and harm-reduction practices are deeply tied to urban planning, demographics, logistics and microeconomics: as scientists we are entrusted to respond to political queries, thus a policy-maker may want to know "how many infections will happen in schools if all grocery stores are closed?", or "what are the kind of workplaces I can close without crashing the economy, which closure would actually benefit public health?". 

Responding soundly to this questions can require quite a bit of mental gymnastics if we're fiddling with differential equations, stochastic processes and statistical mechanics: on the other hand answers come naturally when dealing with multi-agent modeling.

\begin{figure}[htp]

\centering
\includegraphics[width = 0.5\textwidth]{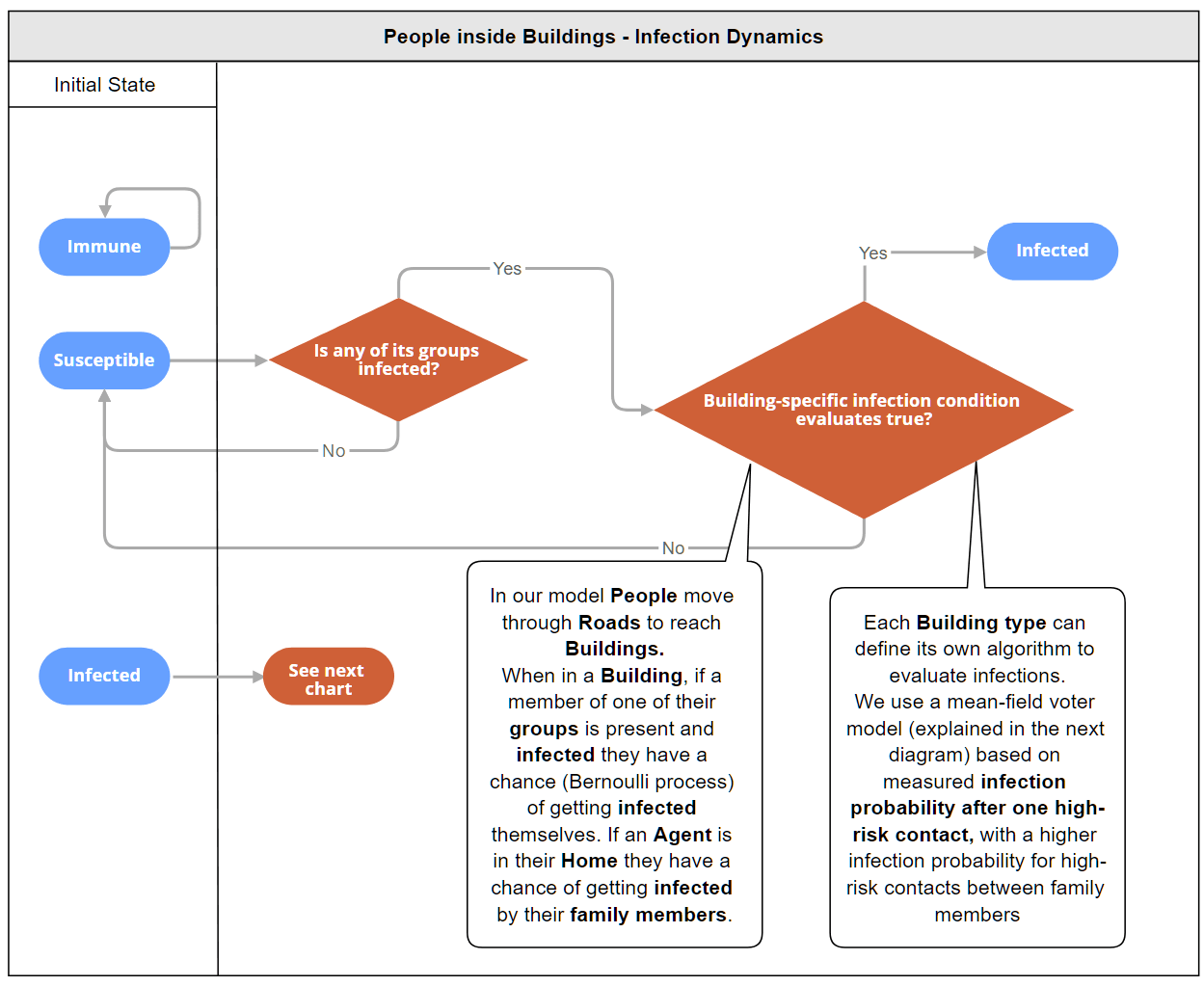}

\caption{Contagion flow-chart for \textit{People}}
\label{fig:PEOPLE_INFECTION}

\end{figure}

\begin{figure}[htp]

\centering
\includegraphics[width = 0.5\textwidth]{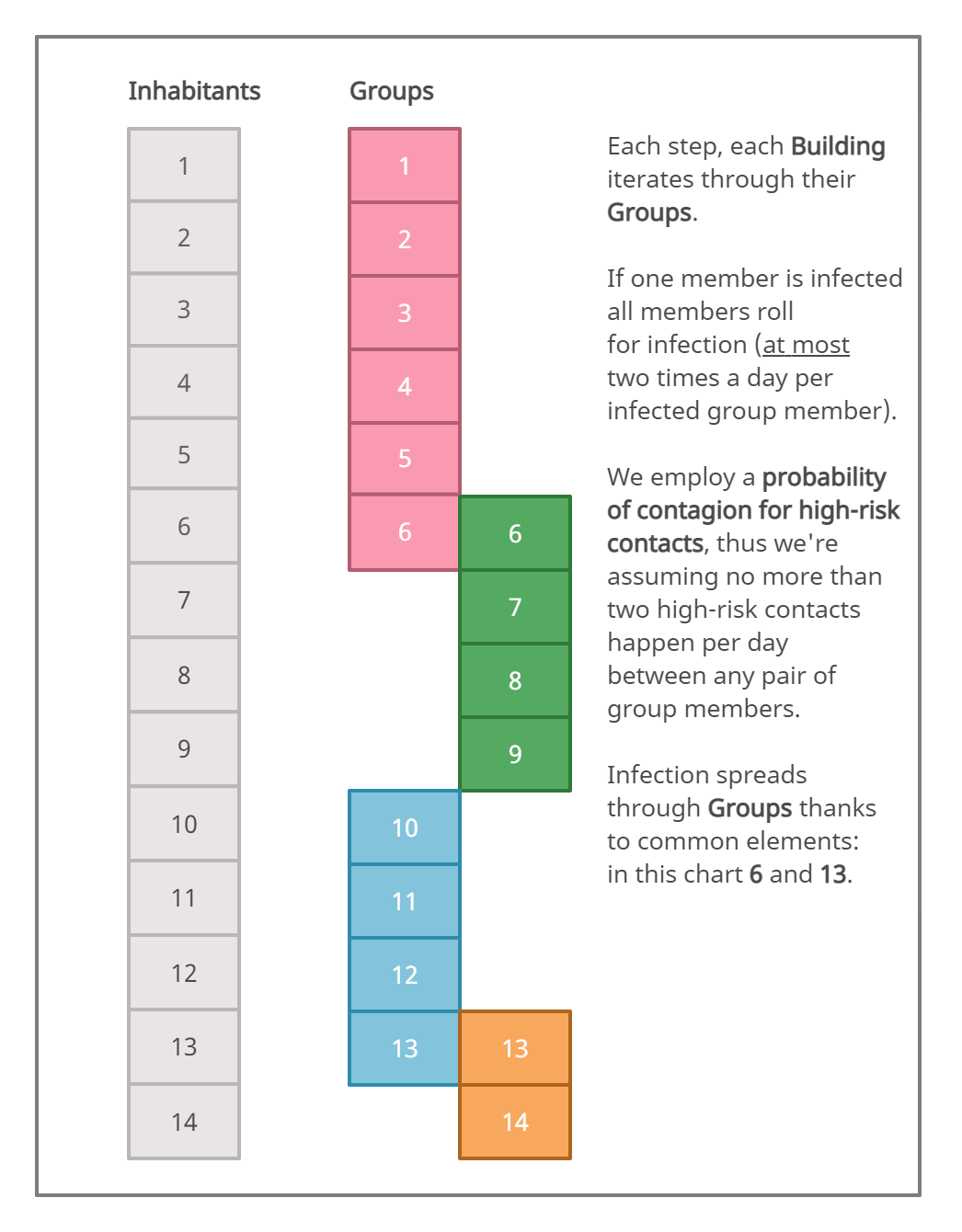}

\caption{Explanatory chart for group contagion in \textit{Buildings}}
\label{fig:BUILDINGS_GROUPS}

\end{figure}

\begin{figure}[htp]

\centering
\includegraphics[width = 0.5\textwidth]{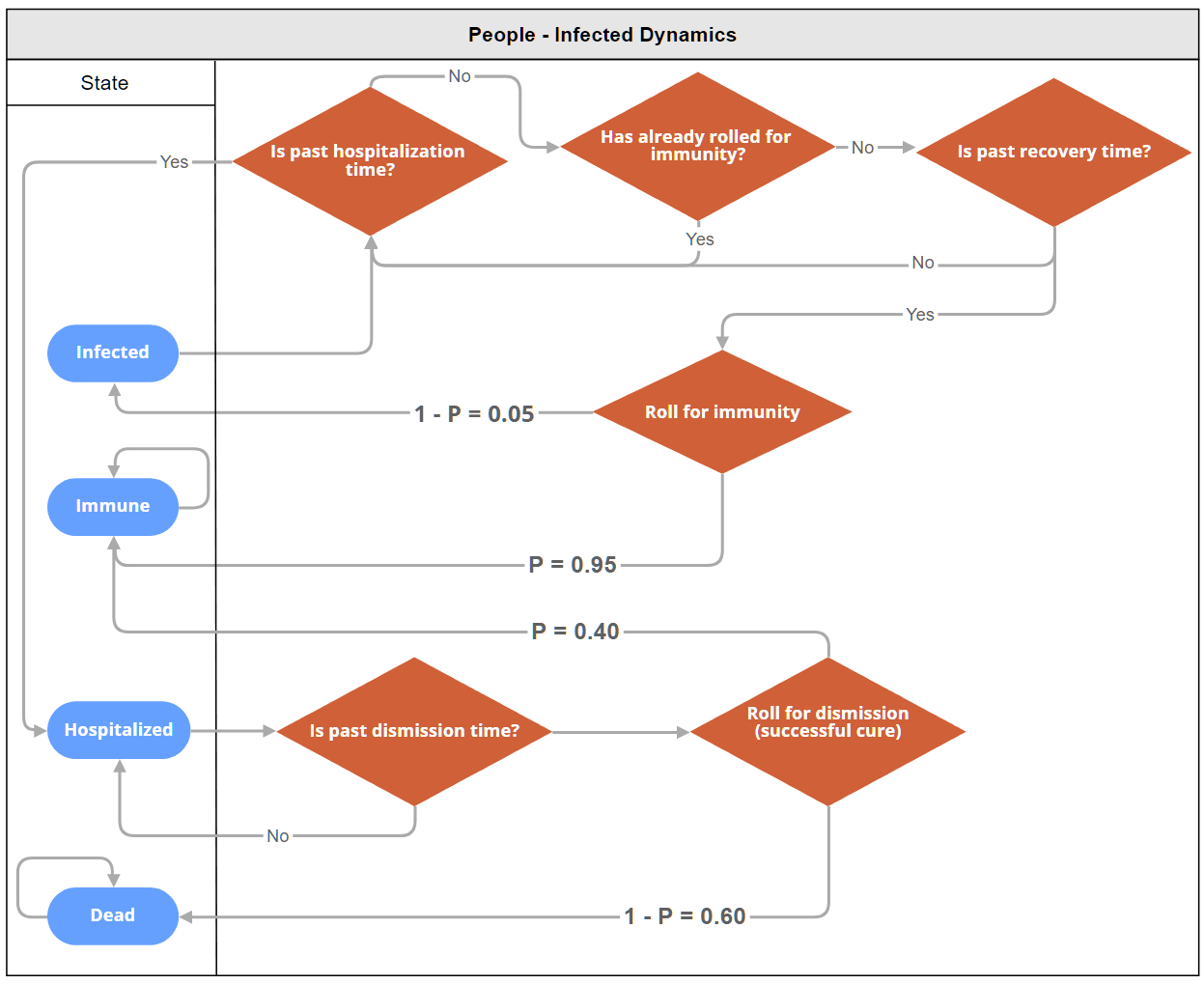}

\caption{Flow-chart for infection development in \textit{People}}
\label{fig:PEOPLE_INFECTED}

\end{figure}

Figures \ref{fig:PEOPLE_INFECTION} and \ref{fig:BUILDINGS_GROUPS} offer a flow-chart for the infection process, acted by \textit{Buildings}.
Figure \ref{fig:PEOPLE_INFECTED} details the evolution of infections in \textit{People}.

\subsection{\textbf{Restriction modeling}}

We model two coexisting classes of restriction:
\begin{itemize}
    \item \textit{Curfew}: a curfew is applied to all agents at a given hour. Working-time is diminished according to the curfew. If caught outside at curfew, agents are compelled to go home. A slight chance remains to take a trip to some shops for food, medicines or any other essential and then head back home. If no lockdown is imposed, all \textit{People} will go to work as usual.
    
    \item \textit{Lockdown}: a fixed proportion (for Toronto it's roughly 85\%) of workers are considered non-essential and will not move to go to work, school or church. They retain their \textit{Resting} schedule and there's a fair chance they will visit stores and parks. If no curfew is imposed \textit{People} will behave as usual in regards to their free time. 
\end{itemize}

We separate these restrictions to control them individually and experiment different lockdown patterns.

Parameter values and restriction details are presented for each simulation in the next section.

\section{\textbf{Simulation Details and Analysis}}

To investigate the peculiarities of our model we opted for two different population sizes: a \textit{realistic case} with $30030$ agents, resembling the actual inhabitants distribution which is roughly $30000$ \footnote{Data have been gathered from \href{https://www.toronto.ca/ext/sdfa/Neighbourhood\%20Profiles/pdf/2016/pdf1/cpa01.pdf}{https://www.toronto.ca/ext/sdfa/Neighbourhood\%20Profiles/pdf/2016/pdf1/cpa01.pdf}}; a \textit{small world case} with $1020$ agents, where we limited the number of buildings and zones to a fraction - approximately $3\%$ - of the factual one.

For both these systems, we also explored two scenarios: the first is an unconstrained epidemics, we tested for different \textit{infection rate} parameter ($\beta$) values ; in the second one, we stuck to a specific $\beta$ and employed restrictions based on Curfew, both by imposing a specific daily time, e.g. 6 p.m., and a delay between the starting of the simulation and the beginning of mitigation measures.

\subsection{\textbf{Small World Case - 1020 citizens}}

\begin{figure}[htp]
\centering
\includegraphics[width = 0.5\textwidth]{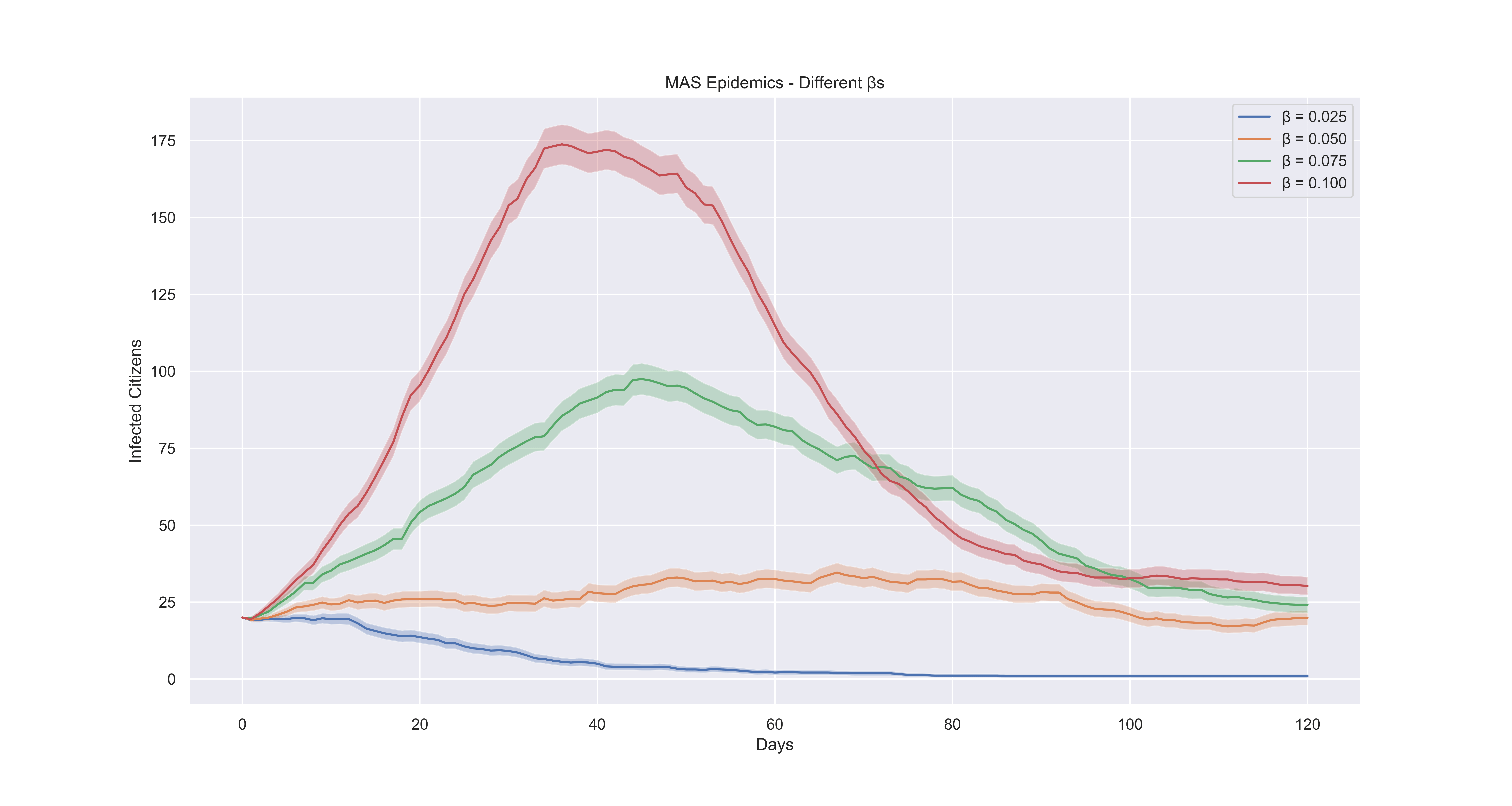}
\caption{Epidemic dynamics for different values of beta parameter in the small world case with 1020 citizens, the shade corresponds to 95\% confidence interval}
\label{fig:betas_1k}
\end{figure}

With this restricted implementation we tested our model evolution for different scenarios and for a time-span compatible with the actual duration of Covid-19's first-wave in the city of Toronto. As in Figure \ref{fig:betas_1k}, we decided to gather information about the dynamics for different values of $\beta$. By exploiting GAMA's \textit{Batch Mode}, we ran and averaged $30$ simulations for each parameter, more details about the peak are in Table \ref{table:I_perc_1020_betas}. 

\begin{table}
  \caption{Percentage of Infected Citizens over 30 Batches at the Peak - 1020 citizens}
  \label{table:I_perc_1020_betas}
   \centering
  \begin{tabular}{llll}
    \toprule
    \multicolumn{2}{c}{Infection Rate}    
     \\
    \cmidrule(r){1-3}
    $\beta$ & Infected \% & 95\% C.I. & Peak Day    
     \\
    \midrule
    0.025 &  N/A & N/A & N/A \\
    0.05 & 3.4 & $\pm$ 0.3 & 67  \\
    0.075 & 9.5  &  $\pm$ 0.5 & 45 \\
    0.1   & 17.0  & $\pm$ 0.6 & 36   \\
  
   \bottomrule
\end{tabular}
\end{table}

\subsection{\textbf{Realistic Case - 30030 citizens}}

\begin{figure}[htp]
\centering
\includegraphics[width = 0.5\textwidth]{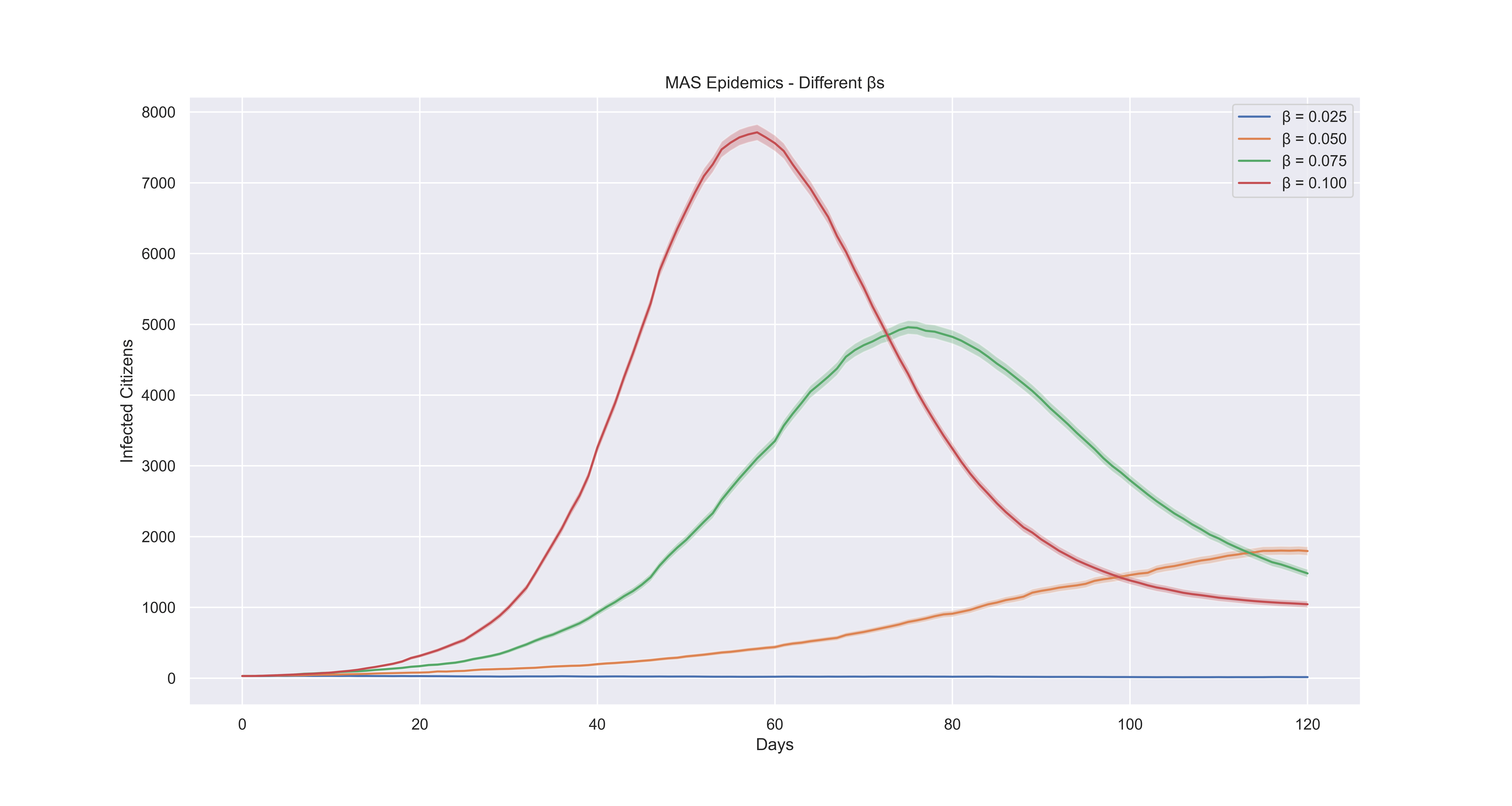}

\caption{Epidemic dynamics for different values of beta parameter in the realistic case with 30030 citizens, the shade corresponds to 95\% confidence interval}
\label{fig:betas_30k}

\end{figure}

As a proof of work, we decided to extend our previously reported toy model to a realistic case, by employing the actual number of citizens. Again, we exploited GAMA's \textit{Batch Mode}, and averaged $30$ simulations for each $\beta$ as in Figure \ref{fig:betas_30k}. Details are also reported in Table \ref{table:I_perc_30030_betas}. 

From a qualitative inspection, this simulation bear resemblance to a more stable evolution, showing SIR-like characteristics. 

\begin{table}
  \caption{Percentage of Infected Citizens over 30 Batches at the Peak - 30030 citizens}
  \label{table:I_perc_30030_betas}
   \centering
  \begin{tabular}{llll}
    \toprule
    \multicolumn{2}{c}{Infection Rate}    
     \\
    \cmidrule(r){1-3}
    $\beta$ & Infected \% & 95\% C.I. & Peak Day    
     \\
    \midrule
    0.025 &  N/A & N/A & N/A \\
    0.05 & 6.0 & $\pm$ 0.2 & 119  \\
    0.075 & 16.5  &  $\pm$ 0.3 & 75 \\
    0.1   & 25.7  & $\pm$ 0.4 & 58   \\
  
   \bottomrule
\end{tabular}
\end{table}

\section{\textbf{RESTRICTIONS ANALYSIS}}

As previously highlighted in paragraph $4.5$, we employed two parallel mitigation measures. Technically speaking these measures map directly into two parameters, set up in GAMA as, namely: \textit{Curfew Time} and \textit{Curfew Delay}. The first one is basically a night-time curfew, starting either from $6 p.m.$ , $7 p.m.$ or $8 p.m.$ in our model. The second one is the specific day after which, from the starting of the simulation, restrictions take place, selected as $5$ days or $10$ days.  

\subsection{\textbf{Small World Case - 1020 citizens}}

\begin{figure}[htp]
\includegraphics[width = 0.5\textwidth]{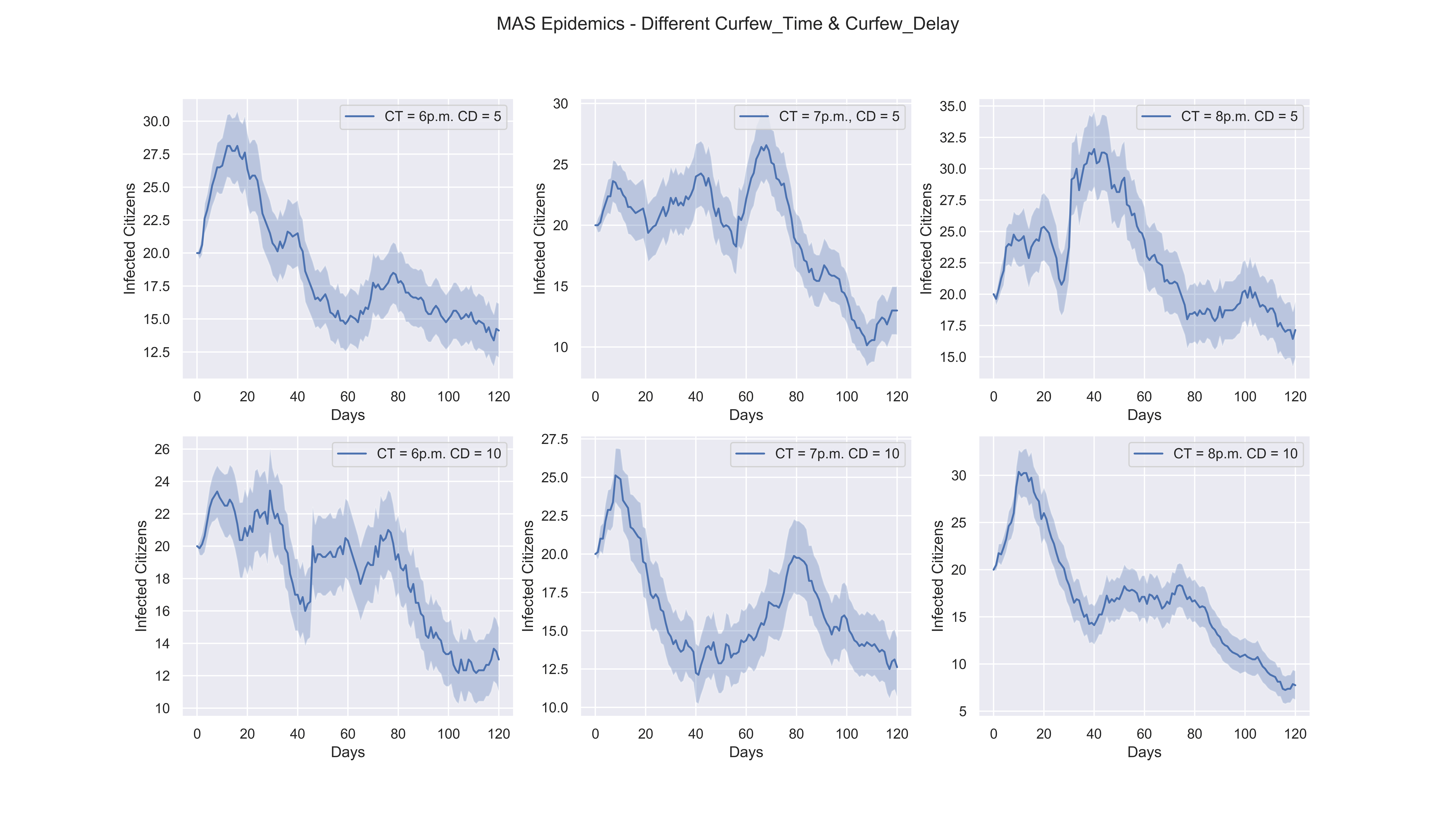}
\caption{Epidemic Dynamics in presence of Curfew for beta = 0.05. CT corresponds to Curfew Time, i.e. the specific hour in the day when curfew starts, instead CD is the Curfew Delay, i.e. how many days pass after the starting of the restrictions. The shade corresponds to 95\% confidence interval}
\label{fig:beta_0.05_1k}
\end{figure}

\begin{figure}[htp]

\includegraphics[width = 0.5\textwidth]{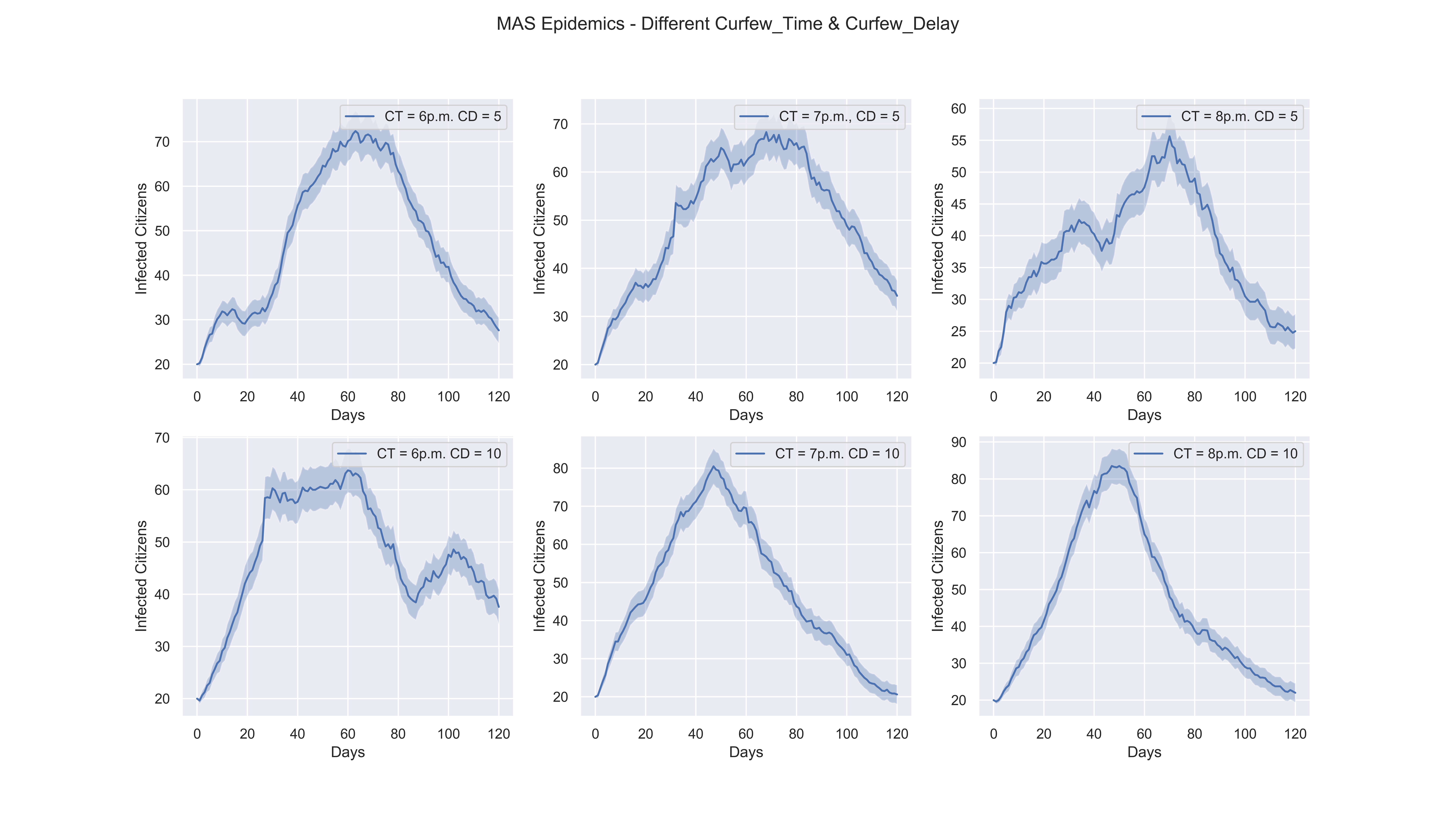}

\caption{Epidemic Dynamics in presence of Curfew for beta = 0.075. The shade corresponds to 95\% confidence interval}
\label{fig:beta_0.075_1k}

\end{figure}

\begin{figure}[htp]

\includegraphics[width = 0.5\textwidth]{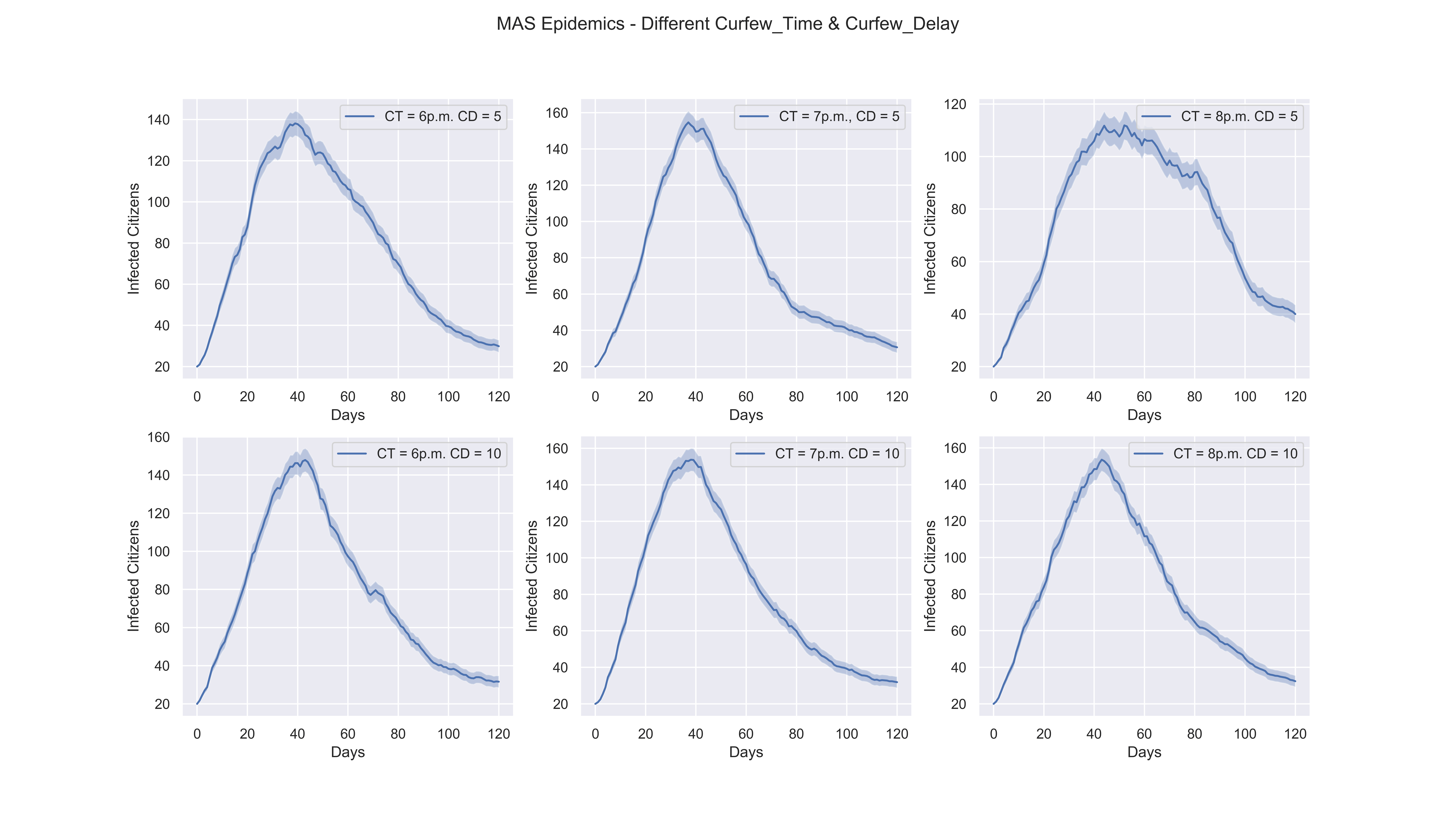}

\caption{Epidemic Dynamics in presence of Curfew for beta = 0.1. The shade corresponds to 95\% confidence interval}
\label{fig:beta_0.1_1k}

\end{figure}

We tested for the combinations between three different values of $\beta$ - namely $0.05$, $0.075$, $0.1$ - , \textit{Curfew Time} and \textit{Curfew Delay} previously stated values as in Figures \ref{fig:beta_0.05_1k}, \ref{fig:beta_0.075_1k} and \ref{fig:beta_0.1_1k}. 

In Tables \ref{table:0.05_curfew_time_delay}, \ref{table:0.075_curfew_time_delay}, \ref{table:0.1_curfew_time_delay} further details are reported. 

Even though this model is a minimalist one and it is certainly less stable with respect to its \textit{realistic} counterpart, some peculiarities can be extracted. For instance, there is a recurrent pattern for different values of $\beta$: the peak is less pronounced if by fixing the \textit{Curfew Delay} at $5$, we postpone \textit{Curfew Time} from $6 p.m.$ to $8 p.m.$ This behaviour might be related to the fact that by closing \textit{commercial} buildings right after the time when most of the people finish to work, they end up being busy and overran by agents: an ideal condition for Covid-19 to thrive. Indeed, by setting \textit{Curfew Time} at $8 p.m.$ we notice a partial relief in this sense. 

\begin{table}
  \caption{$\beta = 0.05$ - Different Tested Parameters and Results for  Small World Case}
  \label{table:0.05_curfew_time_delay}
   \centering
   \resizebox{0.5\textwidth}{!}{
  \begin{tabular}{llllll}
    \toprule
    \multicolumn{2}{c}{Restriction Analysis}                   \\
    \cmidrule(r){1-2}
    $\beta$     & Curfew Time     & Curfew Delay & Infected \% & 95\% C.I. & Peak Day\\
    \midrule
    0.05 & 6 p.m.  & 5  & 2.8 & $\pm$ 0.2 &  12  \\
    0.05 & 6 p.m.  & 10 & 2.3 & $\pm$ 0.3 &  29     \\
    0.05 & 7 p.m.  & 5   & 2.6 & $\pm$ 0.3 &  68   \\
    0.05 & 7 p.m.  & 10   & 2.5 & $\pm$ 0.2 &  8   \\
    0.05 & 8 p.m.  &  5   & 3.0 & $\pm$ 0.3 &  41  \\
    0.05 & 8 p.m.  & 10  & 3.0 & $\pm$ 0.2 &  10   \\
   \bottomrule
\end{tabular}}
\end{table}

\begin{table}
  \caption{$\beta = 0.075$ - Different Tested Parameters and Results for Small World Case}
  \label{table:0.075_curfew_time_delay}
   \centering
   \resizebox{0.5\textwidth}{!}{
  \begin{tabular}{llllll}
    \toprule
    \multicolumn{2}{c}{Restriction Analysis}                   \\
    \cmidrule(r){1-2}
    $\beta$     & Curfew Time     & Curfew Delay & Infected \% & 95\% C.I. & Peak Day\\
    \midrule
    0.075 & 6 p.m.  & 5  & 7.1 & $\pm$ 0.4 & 63 \\
    0.075 & 6 p.m.  & 10 & 6.2 & $\pm$ 0.4 &  60     \\
    0.075 & 7 p.m.  & 5   & 6.8 & $\pm$ 0.4 &  68   \\
    0.075 & 7 p.m.  & 10   & 7.9 & $\pm$ 0.4 &  47   \\
    0.075 & 8 p.m.  &  5   & 5.4 & $\pm$ 0.4 &  70  \\
    0.075 & 8 p.m.  & 10  & 8.1 & $\pm$ 0.5 &  47   \\
   \bottomrule
\end{tabular}}
\end{table}

\begin{table}
  \caption{$\beta = 0.1$ - Different Tested Parameters and Results for Small World Case}
  \label{table:0.1_curfew_time_delay}
   \centering
   \resizebox{0.5\textwidth}{!}{
  \begin{tabular}{llllll}
    \toprule
    \multicolumn{2}{c}{Restriction Analysis}                   \\
    \cmidrule(r){1-2}
    $\beta$     & Curfew Time     & Curfew Delay & Infected \% & 95\% C.I. & Peak Day\\
    \midrule
    0.1 & 6 p.m.  & 5  & 13.5 & $\pm$ 0.6 &  39  \\
    0.1 & 6 p.m.  & 10 &  14.5 & $\pm$ 0.6 &  43     \\
    0.1 & 7 p.m.  & 5   & 15.1 & $\pm$ 0.8 &  37   \\
    0.1 & 7 p.m.  & 10   & 15.1 & $\pm$ 0.6 &  38   \\
    0.1 & 8 p.m.  &  5   & 10.9 & $\pm$ 0.5 &  52  \\
    0.1 & 8 p.m.  & 10  & 15.1 & $\pm$ 0.6 &  43   \\
   \bottomrule
\end{tabular}}
\end{table}

\section{\textbf{CONCLUSIONS}}

The aim of this study was not to model exactly the Toronto epidemic, albeit our model grasps the time-scale and qualitative behavior of the epidemic in West-Humber-Clairville (\textit{circa} 4 months), while overestimating the number of infections. 

We believe further testing of the model should focus on replicating large scale real-world data: the fact that time-scale and qualitative behavior are grasped by this model with minimal assumptions suggests that with more realistic network structures and sample sizes we may pinpoint aggregate epidemic dynamics while maintaining a tool that allows enquiries of much higher significance for policy makers than what is extracted from approaches based on ODEs or SDEs.

Our specific model was developed in GAMA, but the methodological principles from which we have proceeded can be translated to any agent-based modeling framework that enables GIS embedment. Namely:
\begin{itemize}
    \item Choose actors that share open data on zoning laws, contagions, demographics and mobility
    \item Enrich GIS data on the focal area adding zoning information
    \item Enable scheduled (work, school, grocery shopping) and unscheduled (random trips to the store, going to the park, going to church once in a while) movement
    \item The abstract entity representing buildings is the owner of the contagion process
    \item Enable different network structures and risk factors in different building classes: the virus stays the same, our behaviors and relationships differ through each environment we visit.
    \item The abstract entity representing people is the owner of the immunization process
    \item Enable different age-classes that enact characteristic behaviors and react differently to infections
    \item Collect data on the multiplex networks extracted from social networks, layered by building class. Look for static network metrics, topology and structures that influence contagion.
    \item Focus on simulating scenarios that are useful for policy-makers
\end{itemize}

We think some counter-intuitive information is already coming out of our model: as noted in paragraph $6.1$, if we move up the curfew, we'd expect contagions to lower as we raise it towards noon, but this is not the case: it may be best to avoid crowding in essential enterprises than to restrict movement altogether.

\section{\textbf{ACKNOWLEDGEMENTS}}
This research has been initially carried as part of a project for the course \textit{Laboratory on advanced modeling techniques: Multi Agent Systems (MAS)}, taught by Prof. Marco Maggiora at the University of Turin.
We would like to personally acknowledge Jonathan Critchley for useful suggestions and feedbacks with Toronto's geospatial data.
The implementation described in the previous paragraphs is largely based on the Agent-Based framework \textit{GAMA} and it is available on a public Github repository \footnote{Code available at: \href{https://github.com/MachineLearningJournalClub/GAMELEON}{https://github.com/MachineLearningJournalClub/GAMELEON}}. For the development we employed a workstation with the following technical specifications:
\vspace{-5pt}

\begin{itemize}
\itemsep0em
    \item CPU: Intel Core i7 10700K 8 cores / 16 threads
    \item RAM: 64 GB DDR4 unbuffered
    \item GPU: Nvidia 2080 Ti 11 GB GDDR6
\end{itemize}

We deeply acknowledge the Machine Learning Journal Club for providing us with the computational resources, the University of Turin and the University of Ottawa for supporting us.

\clearpage
\printbibliography
\end{document}